\documentclass[a4paper,twocolumn,showpacs,notitlepage,floatfix,nofootinbib,superscriptaddress,pre]{revtex4-1}
\usepackage{graphicx}
\usepackage{amsfonts}
\usepackage{amsmath}
\usepackage{hyperref}
\usepackage{float}
\usepackage{amssymb,amsbsy}
\usepackage{epstopdf}
\usepackage{color}
\usepackage{hhline}
\usepackage{tabularx}
\begin{document}

\title{Direct laser-driven electron acceleration and energy gain in helical beams}
\author{Etele Moln\'ar}
\affiliation{Extreme Light Infrastructure – Nuclear Physics ELI-NP, “Horia Hulubei” 
National Institute for Physics and Nuclear Engineering, 30 Reactorului Street, RO-077125, Bucharest-Magurele, Romania} 

\author{Dan Stutman}
\affiliation{Extreme Light Infrastructure – Nuclear Physics ELI-NP, “Horia Hulubei” 
National Institute for Physics and Nuclear Engineering, 30 Reactorului Street, RO-077125, Bucharest-Magurele, Romania} 

\date{\today }

\begin{abstract}
A detailed study of direct laser-driven electron acceleration in paraxial Laguerre-Gaussian modes corresponding 
to helical beams $\text{LG}_{0m}$ with azimuthal modes $m=\left\{1,2,3,4,5\right\}$ is presented.
Due to the difference between the ponderomotive force of the fundamental Gaussian beam $\text{LG}_{00}$ 
and helical beams $\text{LG}_{0m}$ we found that the optimal beam waist leading to the most energetic electrons
at full width at half maximum is more than twice smaller for the latter and corresponds to a few wavelengths 
$\Delta w_0=\left\{6,11,19\right\}\lambda_0$ for laser powers of $P_0 = \left\{0.1,1,10\right\}$ PW. 
We also found that for azimuthal modes $m\geq 3$ the optimal waist should be smaller than $\Delta w_0 < 19 \lambda_0$.
Using these optimal values we have observed that the average kinetic energy gain 
of electrons is about an order of magnitude larger in helical beams compared to the fundamental Gaussian beam.
This average energy gain increases with the azimuthal index $m$ leading
to collimated electrons of a few $100$ MeV energy in the direction of the laser propagation.
\end{abstract}

\maketitle
\section{Introduction}

The well known Laguerre–Gaussian ($\text{LG}_{pm}$) modes of paraxial light with integer radial and 
azimuthal, $p$ and $m$, indices have a phase distribution 
of $e^{-im\theta}$, where $\theta$ is the azimuthal angle. 
Here $m$ corresponds to the azimuthal or rotational order of the mode 
with a well defined orbital angular momentum (OAM) of $m\hbar$ per photon \cite{Allen_1992}.
Such higher order modes lead to twisted light by forming an optical vortex
in the direction of propagation associated with the optical phase singularity. 
The spiral shape of the wavefront is formed by $|m|$ interconnected helices, hence 
$\text{LG}_{p,m\neq 0}$ beams are called helical beams.

These distinctive properties of higher order Laguerre-Gaussian modes set forth to an ever increasing amount of 
applications in different areas of optical communications, imaging techniques, quantum information technologies 
and other topics, see \cite{Grier_2003, Gibson_2004, Torres_book_2011, Wanget_2012, Krenn_2016} 
and references therein.

In recent years with the development of ultra-intense multi-PW laser technology, Laguerre-Gaussian laser modes have other potential applications in the fields of plasma accelerators, inertial confinement fusion and in the generation 
of X-rays and $\gamma$-rays with OAM \cite{Zhang_2015,Zhang_2016,Ju_2016,Ju_2018,Baumann_2018,Hu_2019,Chen_2019,TundeFulop_2019}.
In particular, it has been shown that in 
laser wakefields driven by $\text{LG}_{01}$ helical pulses, 
the wakefield shows a donut-like structure with a ring-shaped hollow electron beam  \cite{Zhang_2016}. 
Furthermore, for a lower density plasma or a smaller laser spot size, besides the
donut-like wakefield, a central bell-like wakefield forms in the center of 
the donut-like wake. 
On the other hand by further reducing the plasma density or laser spot size leads
to an on-axis electron beam acceleration only. 

It was also demonstrated that Laguerre-Gaussian beams transfer a part of their OAM to electrons 
through the dephasing process similar to the direct electron acceleration (DLA) in Gaussian beams \cite{Nuter_2020}.
Furthermore the propagation of optical beams with OAM leads to plasma waves that may also carry OAM which 
couple to the plasma electrons and involve Landau damping and particle acceleration  
accompanied with generation of quasistatic axial and azimuthal magnetic fields \cite{Blackman_2019}. 
When Laguerre-Gaussian plasma waves are subjected to Landau damping, a higher azimuthal mode number leads 
to a larger OAM transfer to particles traveling close to the phase velocity of the plasma wave \cite{Blackman_2020}.
Vacuum-based charge acceleration with Laguerre-Gaussian beams
has also been studied very recently \cite{Mohammad_2015,Akou_2020} showing that it is possible 
to generate GeV high-quality electron bunch with low spread in energy and radial deflection.

Motivated by these interesting results, 
in this paper we will study direct electron acceleration
in vacuum in various helical Laguerre-Gaussian $\text{LG}_{0m}$ laser pulses corresponding to helical modes 
$m=\left\{ 0,1,2,3,4,5\right\}$. 
Laser beams with $p=0$ and $m=0$ define the fundamental Gaussian mode, while beams with $p=0$ and $m\geq 1$ have a hollow ring like transverse intensity profile with zero intensity at the center.
The central hollowness corresponds to a potential well which confines/accelerates the electric charges through the transverse/longitudinal ponderomotive forces.

This type of ponderomotive trap is also realized using higher order transverse electromagnetic modes (TEM), i.e., the Hermite-Gaussian modes, such as $\text{TEM}_{1,0}$  or the combination of a $\text{TEM}_{1,0}$ with $\text{TEM}_{0,1}$. 
This ponderomotive potential is similarly axisymmetric and has a minimum 
on the axis \cite{Chaloupka_1997,Stupakov_2001,Kawata_2004,Kawata_2005,Kawata_2011}. 
This is certainly expected since orthogonal 
Hermite-Gaussian modes may be decomposed into Hermite-Gaussian modes with a phase difference and vice versa, 
see Refs. \cite{Abramochkin_1991,Allen_1992} for the general formulas.

Here we specifically focus our study on the energy gain in femtosecond lasers of low to very high intensity, 
thereby gaining valuable estimates about relevant parameters for lasers operating at ELI-NP \cite{Tanaka_2020}.
Our results are based on the 3-dimensional numerical 
solution of the relativistic equations of motion for free electrons in paraxial laser fields. 
Henceforth, similarly as in very low density plasmas, in DLA we also observe collimated and intertwining electron 
beams in the direction of the laser propagation, while in addition we also show that the average energy gain 
increases with the azimuthal mode index $m$.

Furthermore, we found that the optimal beam waist leading to the most energetic electrons 
is more than twice smaller in case of higher order Laguerre-Gaussian beams than in case of the fundamental 
Gaussian beam that was presented in \cite{Molnar_2020}. 
For lasers of $P_0 = \left\{0.1,1,10\right\}$ PW power, the optimal beam waists are 
$\Delta w_0=\left\{6,11,19\right\}\lambda_0$ and correspond to 
$I_0 = I^{0m}\times\left\{0.76, 2.2 ,7.6\right\}10^{21}\textrm{W}/\textrm{cm}^2$ peak 
intensities, while taking into account the intensity profiles of the $\text{LG}_{0m}$ beams these values 
are reduced by $I^{0m} \approx \left\{0.37, 0.27, 0.23, 0.19, 0.17\right\}$ 
for $m=\left\{1,2,3,4,5\right\}$.
Using these optimal values we have observed that for the same laser power the average kinetic energy gain 
of electrons is about an order of magnitude larger in helical beams compared to the fundamental Gaussian beam.

The paper is organized as follows. In Sect. \ref{LG-beams} we present 
some of the characteristic properties of helical beams, the equations of motion 
for electrons and the initial conditions corresponding to our study.
In Sect. \ref{Optimal_waist} we have estimated the optimal values of beam waist leading 
to the most energetic electrons for given laser power.
Applying these optimal values we present and discuss the electron dynamics and 
energy gains in linearly polarized (LP) and circularly polarized (CP) helical beams 
Sect. \ref{Energy_gain}. The conclusions are summarized in Sect. \ref{Conclusions}.

\section{Direct laser-driven electron acceleration in helical beams}
\label{LG-beams}
\subsection{Laguerre-Gaussian beams}

A well known solution \cite{Siegman_Lasers_1986,Goldsmith_book_1998} 
to the paraxial wave equation is obtained in
cylindrical coordinates $\left( r,\theta ,z\right) $, with the help of
generalized Laguerre polynomials. These solutions are cylindrically
symmetric around the axis of propagation $z$, with radius 
$r=\sqrt{x^{2}+y^{2}}$ and azimuth $\theta =\arctan\left( x/y\right)$, 
expressed in Cartesian coordinates. 
These are the Laguerre-Gaussian ($\text{LG}_{pm}$) beams 
\cite{Allen_1992, Baumann_2018,Chen_2019}, with radial index $p$ and azimuthal index $m$. 
The general expression for the electric field distribution of a monochromatic $\text{LG}_{pm}$ pulse is,
\begin{align} \label{E_Laguerre} 
& E_{T,pm}\left( r,\theta,z\right) 
= C_{pm} E_{0}\exp \left[-ik_{0}z+i\phi _{0}\right]  \notag \\
& \times
\frac{w_{0}}{w\left( z\right) }\exp \left[ -\frac{r^{2}}{%
w^{2}\left( z\right) }-i\frac{zr^{2}}{Z_{R}w^{2}\left( z\right) }
+i\arctan\left( \frac{z}{Z_{R}}\right) \right]  \notag \\
&\times \left( \frac{\sqrt{2}r}{w\left( z\right) }\right)
^{\left\vert m\right\vert }  \exp \left[ i\left( 2p+\left\vert m\right\vert \right) 
\arctan\left( \frac{z}{Z_{R}}\right) \right]  \notag \\
& \times  L_{p}^{\left\vert m\right\vert }\left( \frac{%
2r^{2}}{w^{2}\left( z\right) }\right) \exp \left[  -im\theta \right],
\end{align}%
where $k_0=\omega_0/c$ is the wavenumber, $c=1/\sqrt{\epsilon_{0}\mu_{0}}$ 
is the speed of light in vacuum, $\omega_0$ is the angular frequency, 
$E_{0}$ is the amplitude of the electric field, and $\phi_{0}$ is the initial phase.
Furthermore, $Z_{R}= w_{0}^{2}k_{0}/2$ is the Rayleigh range, $R_c(z) = \left(z^{2}+Z_{R}^{2}\right) /z $ 
is the radius of curvature, and $\phi _{G}\left( z\right) = \arctan\left( z/Z_{R}\right) $ is 
the Gouy phase.
The beam waist is defined as,
\begin{equation}
w\left( z\right) = w_{0}\sqrt{1+\left( \frac{z}{Z_{R}}\right) ^{2}},
\end{equation}%
where the beam waist radius at focus is $w_{0}\equiv w(z=0)$.

The Laguerre polynomials are denoted by $L_{p}^{m}$, and the normalization constant, 
$C_{pm}=\sqrt{\frac{p!}{\left( p+\left\vert m\right\vert \right) !}}$, follows from the 
orthonormality of the Laguerre polynomials \cite{Gonzalez_2018}.
The fundamental Gaussian beam $\textrm{LG}_{00}$ is obtained for $p = m \equiv 0$, 
where $C_{00} = L_{0}^{0}\equiv 1$.
 
The components of the electric and magnetic fields are
\begin{eqnarray} \label{ExEy_Laguerre} 
E_{x,pm}&=& \alpha_{x} E_{T,pm}, \quad E_{y,pm} = -i \alpha_{y} E_{T,pm},  \\
B_{x,pm} &=&-\frac{1}{c}E_{y,pm} ,\quad B_{y,pm} = +\frac{1}{c}E_{x,pm} .
\label{BxBy_Laguerre}
\end{eqnarray}
where $\alpha_{x}=\sqrt{\left( 1+\alpha _{P}\right) /2}$ and 
$\alpha_{y}=\sqrt{\left( 1-\alpha _{P}\right) /2}$ such that $\alpha _{P}=1$ 
or $-1$ in case of linear polarization along the $x$-axis or $y$-axis respectively, 
and $\alpha _{P}=0$ for circular polarization, while elliptic polarization
otherwise.

The longitudinal components of the electric and magnetic fields are
calculated from Maxwell's equations, 
$\nabla \cdot \vec{E}=\nabla \cdot \vec{B} \equiv 0$, thus in the paraxial approximation,
\begin{eqnarray} \nonumber
E_{z,pm} &\equiv &-\frac{i}{k_{0}}\left( \frac{\partial E_{x,pm}}{\partial x}
+\frac{\partial E_{y,pm} }{\partial y}\right) \\
&=& \frac{i}{k_{0} w^{2}\left( z\right) } \left[ x E_{x,pm}+ y E_{y,pm}\right] \nonumber \\ 
&\times &  \left[ 2\left(1 + i \frac{z}{Z_R} \right) - \left\vert m\right\vert \frac{w^{2}\left( z\right) }{r^{2}} 
+ \frac{4 L_{p-1}^{\left\vert m\right\vert +1}\left( \frac{2r^{2}}{w^{2}\left(
z\right) }\right) }{L_{p}^{\left\vert m\right\vert }\left( \frac{2r^{2}}{%
w^{2}\left( z\right) }\right) }\right] \nonumber \\
&-& \frac{m}{k_{0}r^{2}}\left[ y E_{x,pm} - x E_{y,pm}\right],  \label{Ez_Laguerre} \\
B_{z,pm} &\equiv &-\frac{i}{ck_{0}}
\left( \frac{\partial E_{x,pm}}{\partial y} - \frac{\partial E_{y,pm}}{\partial x}\right) \nonumber \\
&=& \frac{i}{c k_{0} w^{2}\left( z\right) } \left[ y E_{x,pm} - x E_{y,pm}\right] \nonumber \\ 
&\times &  \left[ 2\left(1 + i \frac{z}{Z_R} \right) - \left\vert m\right\vert \frac{w^{2}\left( z\right) }{r^{2}} 
+ \frac{4 L_{p-1}^{\left\vert m\right\vert +1}\left( \frac{2r^{2}}{w^{2}\left(
z\right) }\right) }{L_{p}^{\left\vert m\right\vert }\left( \frac{2r^{2}}{%
w^{2}\left( z\right) }\right) }\right] \nonumber \\
&+& \frac{m}{c k_{0}r^{2}}\left[ x E_{x,pm} + y E_{y,pm}\right].  \label{Bz_Laguerre}
\end{eqnarray}

Therefore the electromagnetic field of Laguerre-Gaussian pulses are given by 
\begin{eqnarray}
\vec{E}_{pm}\left( t,r,\theta,z\right) &=& 
\text{Re}\left[ \vec{E}_{pm}\left(r,\theta,z\right) g\left( t,z\right) \right] ,  \label{E_full} \\
\vec{B}_{pm}\left( t,r,\theta,z\right) &=& 
\text{Re}\left[ \vec{B}_{pm}\left(r,\theta,z\right) g\left( t,z\right) \right] ,  \label{B_full}
\end{eqnarray}
where the Gaussian temporal envelope  with $\tau_{0}$ duration, 
and peak intensity position at $z_{F}$ reads  \cite{Arefiev_2013}
\begin{equation}
g\left(t,z\right) = \exp \left[ i\omega _{0}t-\left( \frac{t-\left(
z-z_{F}\right) /c}{\tau _{0}}\right) ^{2}\right] .  \label{time_envelope}
\end{equation}%
Note that there are more appropriate choices for the temporal profile
such as the hyperbolic secant 
$g\left(t,z\right) \sim 1/\cosh \left( \frac{t-\left(z-z_{F}\right) /c}{\tau _{0}}\right)$, see 
Refs. \cite{McDonald_1,McDonald_2,Ong_2018} for more details. 
Using this profile we found that the energy gain may be reduced by as much as $30 \%$ compared to 
the Gaussian.

\subsection{Helical beams}

The Laguerre-Gaussian beams, $\text{LG}_{0m}$, with non-zero azimuthal modes 
$\left\vert m\right\vert\neq 0$ contain a phase change given by $e^{-i m\theta}$.
Note however that for all modes the generalized 
Laguerre polynomials have a contribution equal to one, i.e., $L_{0}^{m}(x) = L_{0}(x)=1$. 
These type of Laguerre-Gaussian beams, $\text{LG}_{0m}$, are the helical beams 
associated with the nonzero OAM of light \cite{Allen_1992,Longman_2020}.

\begin{figure}[hbt!]
\vspace{-0.2cm} 
\includegraphics[width=7.8cm, height=5.6cm]
{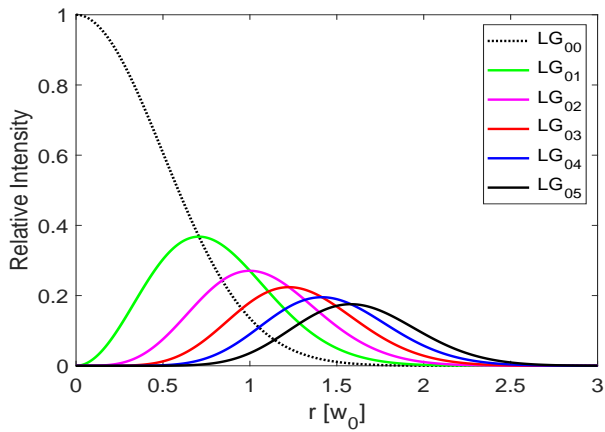}
\caption{The relative intensity profiles at constant power and beam waist
as a function of the radius of an LP laser: The fundamental Gaussian beam $\text{LG}_{00}$
with dotted black line and the helical beams $\text{LG}_{01}$, $\text{LG}_{02}$, $\text{LG}_{03}$,  
$\text{LG}_{04}$ and  $\text{LG}_{05}$ with green, magenta, red, blue 
and black lines correspondingly.}
\label{fig:LG0x_profiles}
\end{figure}

In Fig. \ref{fig:LG0x_profiles} the normalized intensity profiles of a linearly polarized 
fundamental Gaussian and different helical beams are shown as a function of radial distance, 
in units of beam waist radius $w_0$. These plots represent $I^{0m} = |E^{0m}_x/E_0|^2$
using Eq. (\ref{E_Laguerre}), for $t=z\equiv 0$, $k_0 = \omega_0 = w_0 \equiv 1$, 
while the intensity is,  $I^{0m}_0 \sim  E^2_0 I^{0m}$. 

The intensity of $\text{LG}_{0m}$-modes is largest for $m=0$ corresponding to the fundamental
Gaussian beam, and it is decreasing with increasing number of azimuthal modes.
Due to the $\left( \frac{\sqrt{2}r}{w\left( z\right) }\right)^{\left\vert m\right\vert }$ factor,
all modes with $\left\vert m\right\vert \ge 1$ have zero intensity at the center, 
$r = 0$, that is known as an optical vortex or phase singularity on the axis.

The intensity profile of Gaussian beams are concave functions on the whole interval. 
The intensity of helical beams is independent of $\theta$, and 
has a convex part with maxima at $r_m = w_0 \sqrt{|m|/2}$ after 
which the functions change from convex to concave. 
The width of the convex part widens with increasing azimuthal index $m$,
while using the positions of maxima the intensity peaks of $\text{LG}_{0m}$-modes relative to the 
fundamental Gaussian leads,
$I^{0m}(r_m) = C_{0m}{\left\vert m \right\vert}^{\left\vert m \right\vert} e^{-\left\vert m\right\vert }$,
as shown in Fig. \ref{fig:LG0x_profiles}.

\subsection{Electron acceleration in electromagnetic field}

The motion of electrically charged particles in an external electromagnetic field is 
governed by the Lorentz force, 
$\vec{F}_L \equiv d\vec{p}/dt = q \left(\vec{E} + \vec{v}\times \vec{B}\right)$, 
and leads the following set of non-linear differential equations \cite{Jackson_Book_1999},
\begin{eqnarray}
\frac{d\vec{x}}{dt} &=&c\vec{\beta},  \label{dx_dt} \\
\frac{d\vec{\beta}}{dt} &=&\frac{-e}{\gamma m_{e}c}\left[ -\vec{\beta}\left( 
\vec{\beta}\cdot \vec{E}\right) + \vec{E}+c\vec{\beta}\times \vec{B}\right] \, ,
\label{dbeta_dt}
\end{eqnarray}
Here $q=-e$ is the electron's charge, the Cartesian coordinates and normalized velocity 
are denoted by $\vec{x}$ and $\vec{\beta}\equiv \vec{v}/c$, while the Lorentz factor is 
$\gamma = 1/\sqrt{1- |\vec{\beta}|^2}$.

The four momentum of the electron is $p^{\mu }\equiv \left( p^{0},\vec{p}\right)
=m_{e}\gamma c\left( 1,\vec{\beta}\right) $ where 
$m_{e}\equiv\sqrt{p^{\mu}p_{\mu }/c^{2}} =0.511 \, \textrm{MeV}/c^2$ is its invariant rest mass. 
The relativistic energy $\mathcal{E}$ and momentum $\vec{p}$ of electrons are expressed as
\begin{equation}
p^{0}\equiv \frac{\mathcal{E}}{c}=\gamma m_{e}c,\quad \vec{p}=\gamma m_{e}\vec{v}.
\end{equation}%

Laser-driven electron acceleration in vacuum is the consequence of the
direct interaction of the laser pulse with electrons 
\cite{Hartemann_1995,Hartemann_1998,Esarey_1995,Quesnel_1998,Salamin_2002,Salamin_2006,Gupta_2007,Zhang_2008,Fortin_2010,He_2011}.
At any given time, Eqs. (\ref{dx_dt},\ref{dbeta_dt}), 
are input for the electromagnetic field of the laser pulse, i.e., Eqs. (\ref{E_full}, \ref{B_full}), 
hence the trajectory of the propagating electric charge dynamically maps
the laser pulse. 

The 3-dimensional solutions to the electron trajectories and velocities are
obtained by solving these coupled differential equations numerically 
by an adaptive time-step Runge-Kutta method with an accuracy and numerical precision up to 12-digits. 

\begin{figure*}[hbt!]
\vspace{-0.2cm} 
\includegraphics[width=5.6cm, height=4.6cm]
{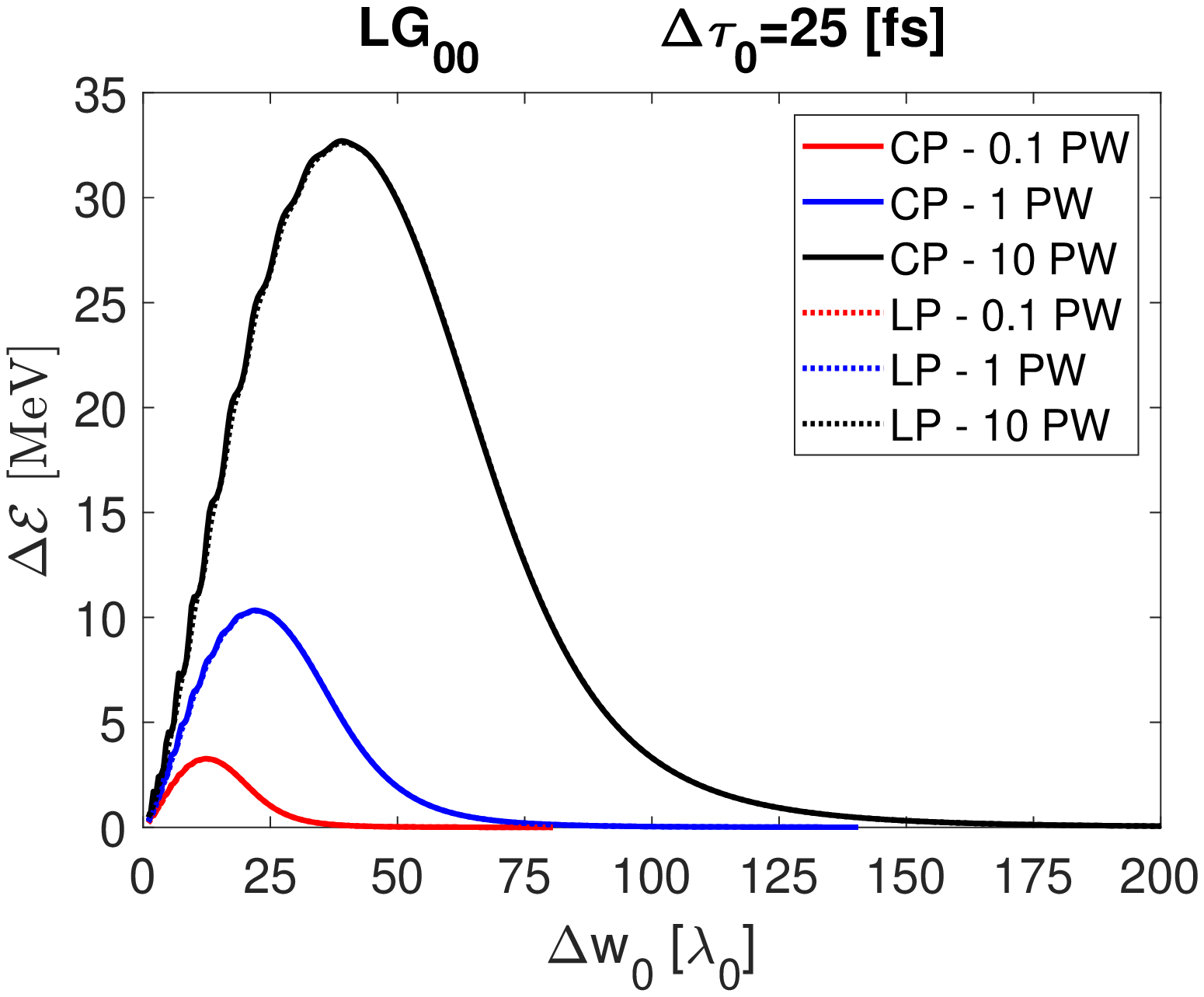} 
\includegraphics[width=5.6cm, height=4.6cm]
{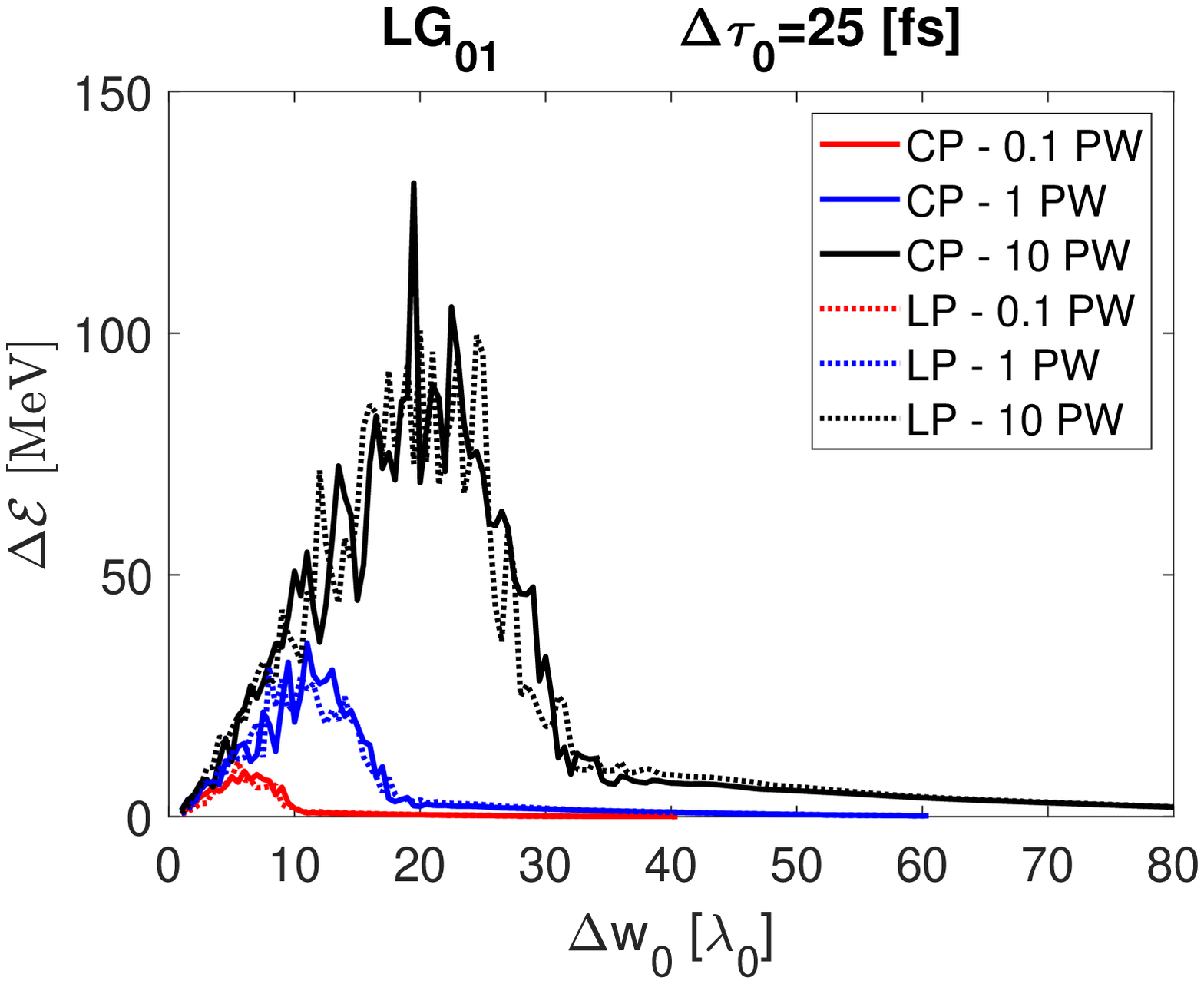} 
\includegraphics[width=5.6cm, height=4.6cm]
{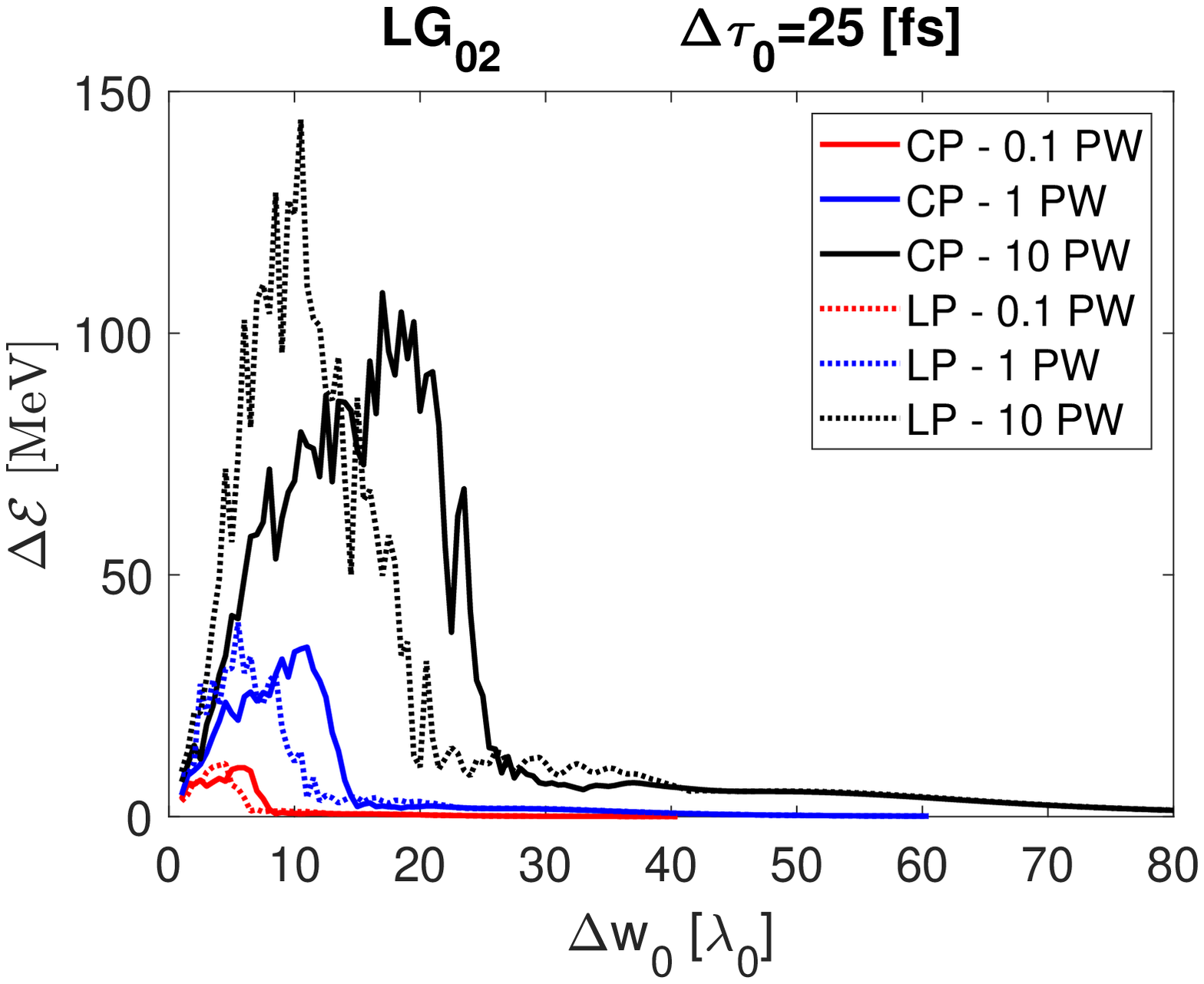} 
\includegraphics[width=5.6cm, height=4.6cm]
{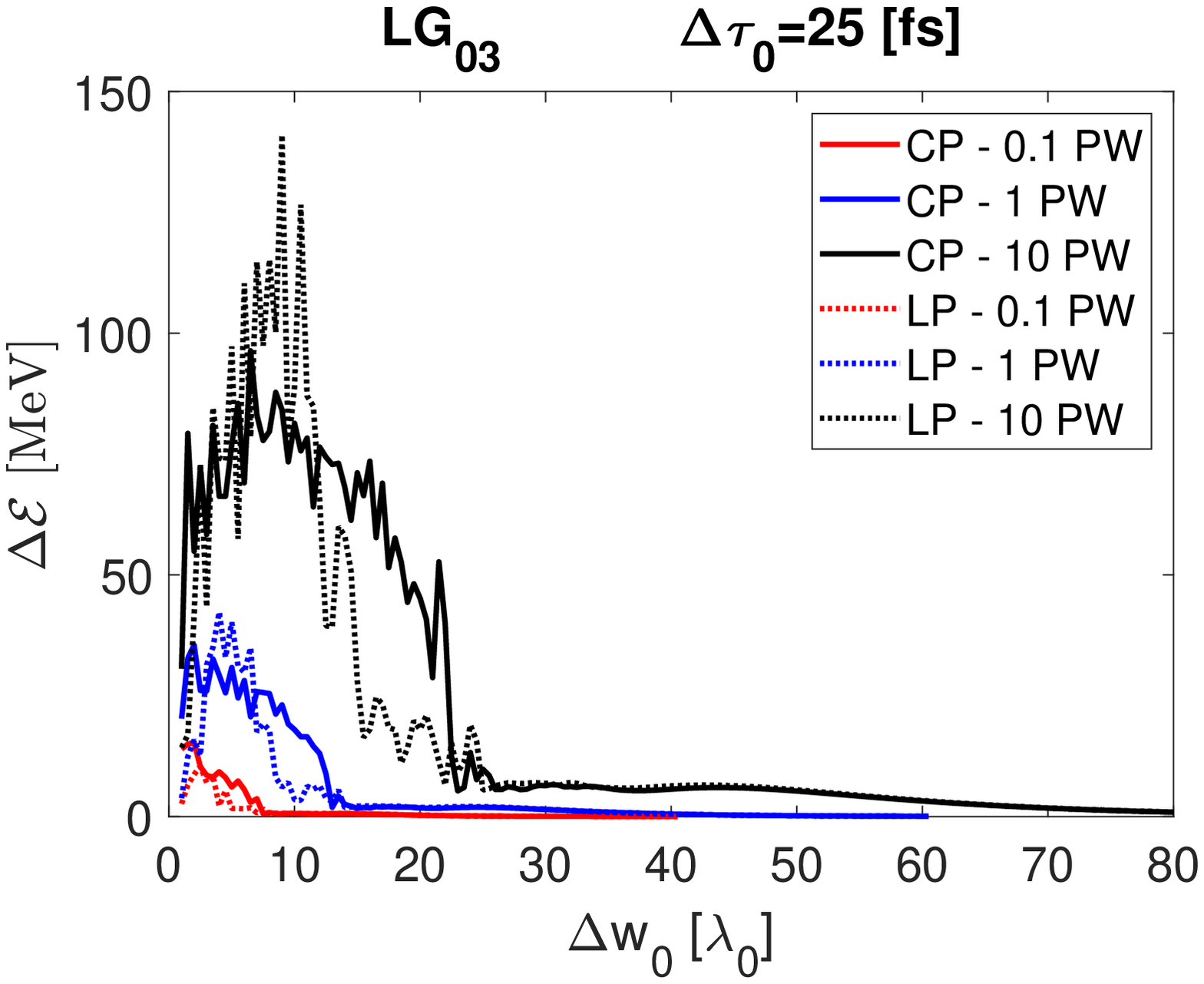}
\includegraphics[width=5.6cm, height=4.6cm]
{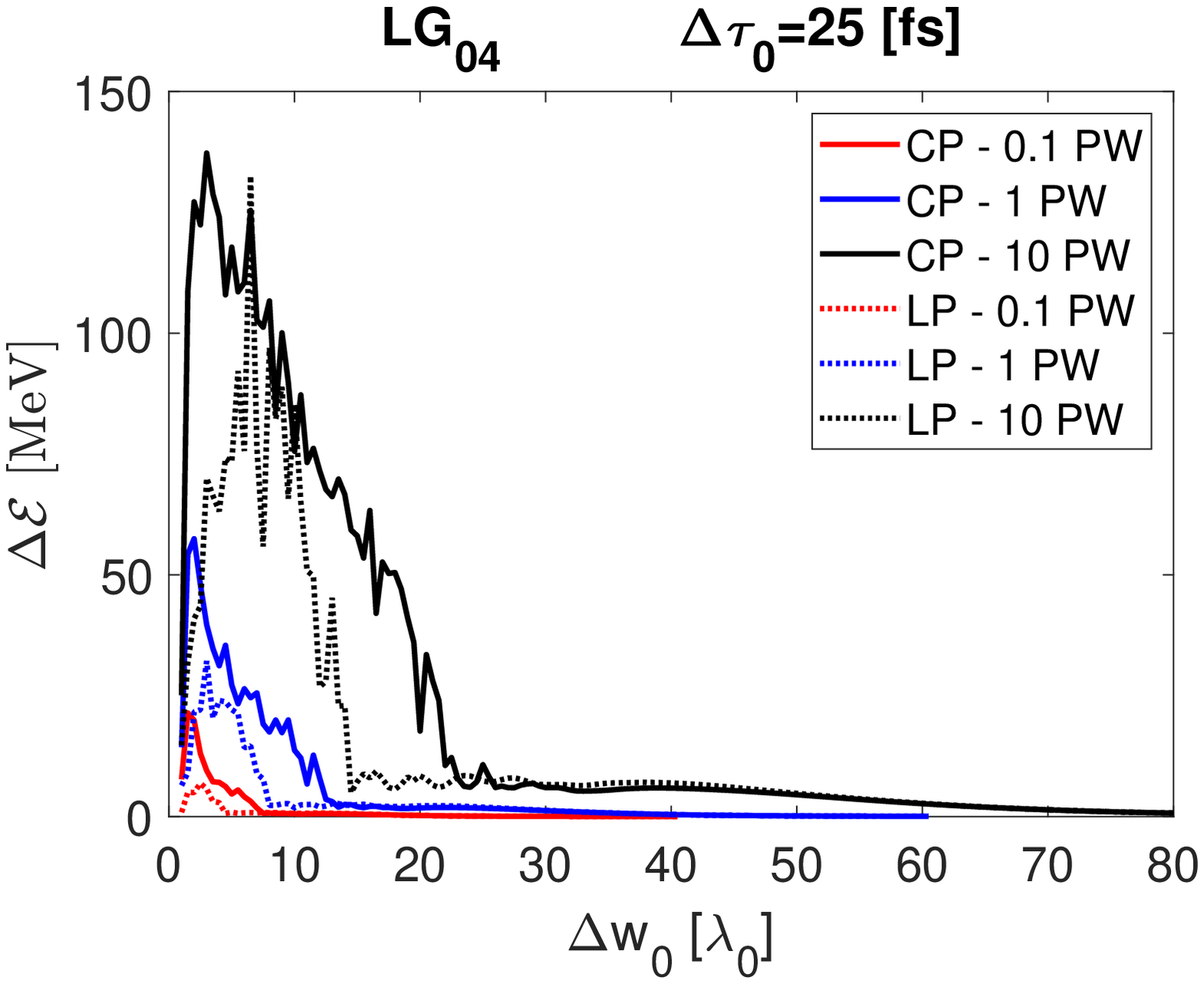}
\includegraphics[width=5.6cm, height=4.6cm]
{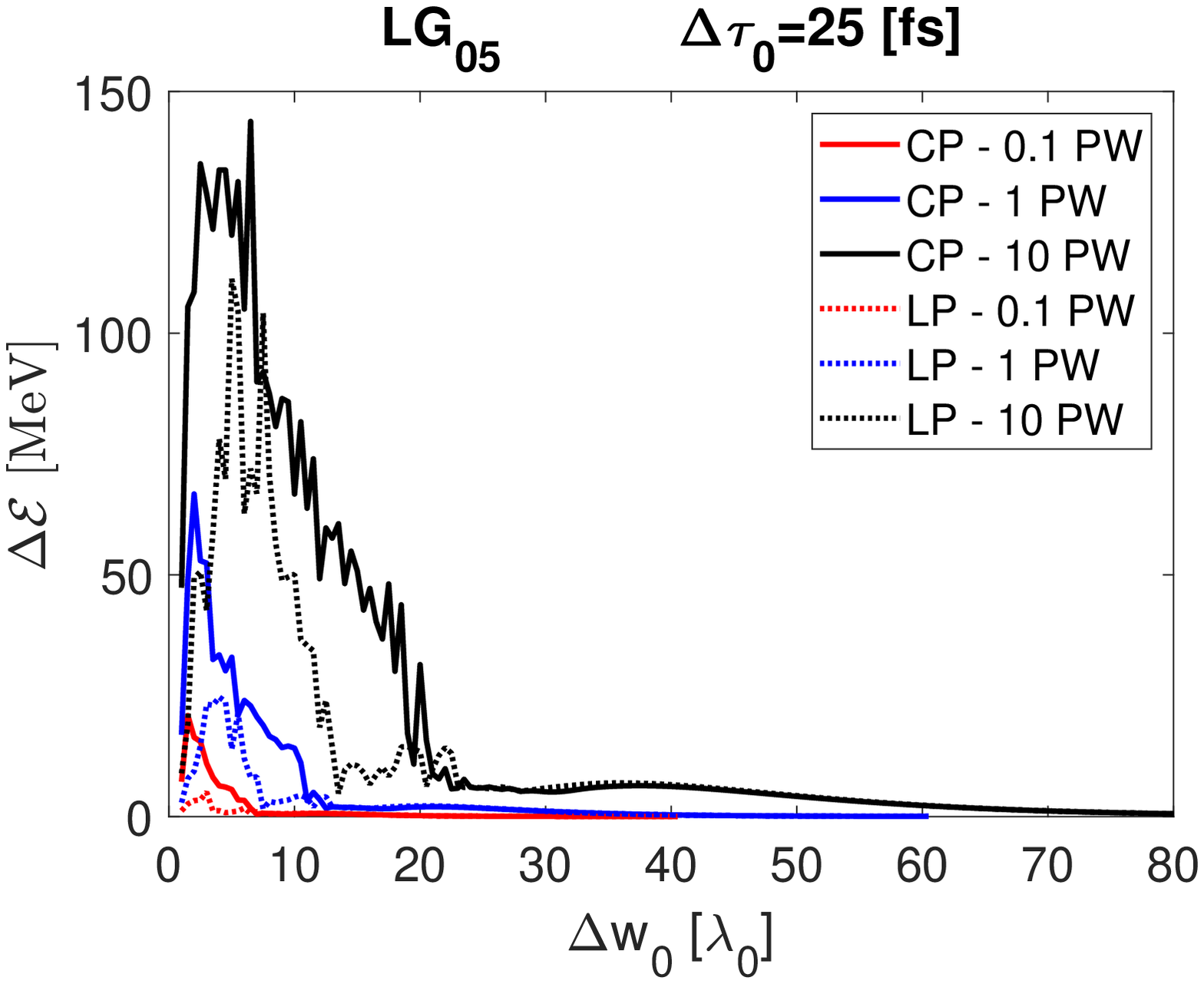}
\caption{The average energy gain of an electron
initially located at $z_0 =0$ and $x_0=y_0 \equiv \left\{0.05,0.1,\cdots,2.5 \right\} \Delta w_0$ 
as a function of the beam waist. 
The red, blue and black lines correspond to the weighted average energy
of $P_0=\left\{0.1, 1, 10\right\}$ PW  laser powers respectively. 
The full and dotted lines correspond to CP and LP lasers. 
The 1st row, from left to right: The Gaussian beam, $\text{LG}_{00}$, 
and the helical beams $\text{LG}_{01}$ and  $\text{LG}_{02}$. 
The 2nd row, from left to right: the helical beams $\text{LG}_{03}$, $\text{LG}_{04}$ and $\text{LG}_{05}$.}
\label{fig:w0_scan}
\end{figure*}

\subsection{Initial conditions}
\label{initial_conditions}

Unless stated otherwise, initially all electrons are at rest, i.e., $\overrightarrow{\beta}_{i} (t_0) = \overrightarrow{\beta}_{0,i} \equiv\overrightarrow{0}$ 
and $\gamma_{0,i} =1$.
These electrons at $z_{0,i}=0$ coordinates are uniformly distributed in the orthogonal plane, 
$\left(x_{0,i},y_{0,i}\right)$, 
on a disk with a radius that is three times the beam waist radius, i.e., $r_{0}=3w_{0}$. 
Thus, initially over 99\% of the laser's energy is contained within this disk.
The initial position of the peak of the laser pulse 
is located on the longitudinal axis at $z_{F}=-5\tau _{0}c$ behind the electrons, i.e., full-pulse interaction, 
while all electrons are independent from each other and only interact with the laser pulse \cite{Molnar_2020}.

For current purposes we have fixed the laser wavelength to 
$\lambda_{0}=800$ nm. The laser pulse duration at Full Width at Half Maximum (FWHM), 
$\Delta \tau _{0}=25$ fs, corresponds to 
$\tau _{0}= \Delta \tau _{0}/2\sqrt{\ln 2}$.
Similarly the beam waist at
FWHM $\Delta w_{0}$ leads to $w_{0}=\Delta w_{0}/2\sqrt{\ln 2}$ waist radius.

The peak intensity and peak power for a monochromatic LP Gaussian laser are 
$I_{0}\equiv a_{0}^{2}\left( \frac{m_{e}c \omega _{0}}{e}\right)^{2}\frac{c\varepsilon _{0}}{2} $ and 
$P_0\equiv I_{0}\frac{\pi w_{0}^{2}}{2}$,
where the normalized electric field amplitude is 
$a_{0}=\frac{eE_{0}}{m_{e}c\omega _{0}}$.
For $P_0=\left\{ 0.1,1,10\right\}$ PW power
and beam waists of $\Delta w_{0}=\left\{ 6,11,19\right\}\lambda_{0}$, 
these values are listed in Table. \ref{Power_table}.
For helical beams the field intensities 
at the local maxim are 
$a_0^{0m}(r_m) = a_0 \sqrt{C_{0m}{\left\vert m \right\vert}^{\left\vert m \right\vert} e^{-\left\vert m\right\vert }}$,
since the total power of the beam is constant, 
$P_0 \equiv \frac{E^2_0}{m!} \int_0^{\infty}  2 \pi r \left( \sqrt{2}r/w_0\right)
^{2\left\vert m\right\vert } \exp \left( -2r^{2}/w^{2}_0 \right) dr = E^2_0  w^2_0\pi/2 $. 

\begin{table}[hbt]
\centering
\begin{tabular}{|c|c|c|c|}
\hline
$P_0 \,[\textrm{PW}]$ & $w_0=3.6\lambda_0$ & $w_0=6.6\lambda_0$ & $w_0=11.4\lambda_0$ 
\\  \hhline{=|=|=|=|}
$0.1$ & $a_0=18.9$ & $a_0=10.3$ & $a_0=5.9$ \\ \hline
$1$   & $a_0=59.8$ & $a_0=32.6$ & $a_0=18.9$ \\ \hline
$10$  & $a_0=189.3$ & $a_0=103.2$ & $a_0=59.7$ \\ \hline
\end{tabular}
\caption{The normalized field amplitudes corresponding
to LP Gaussian pulses for different waist radii and laser powers.}
\label{Power_table}
\end{table}

\begin{figure*}[hbt!]
\vspace{-0.2cm} 
\includegraphics[width=17.2cm, height=4.2cm]
{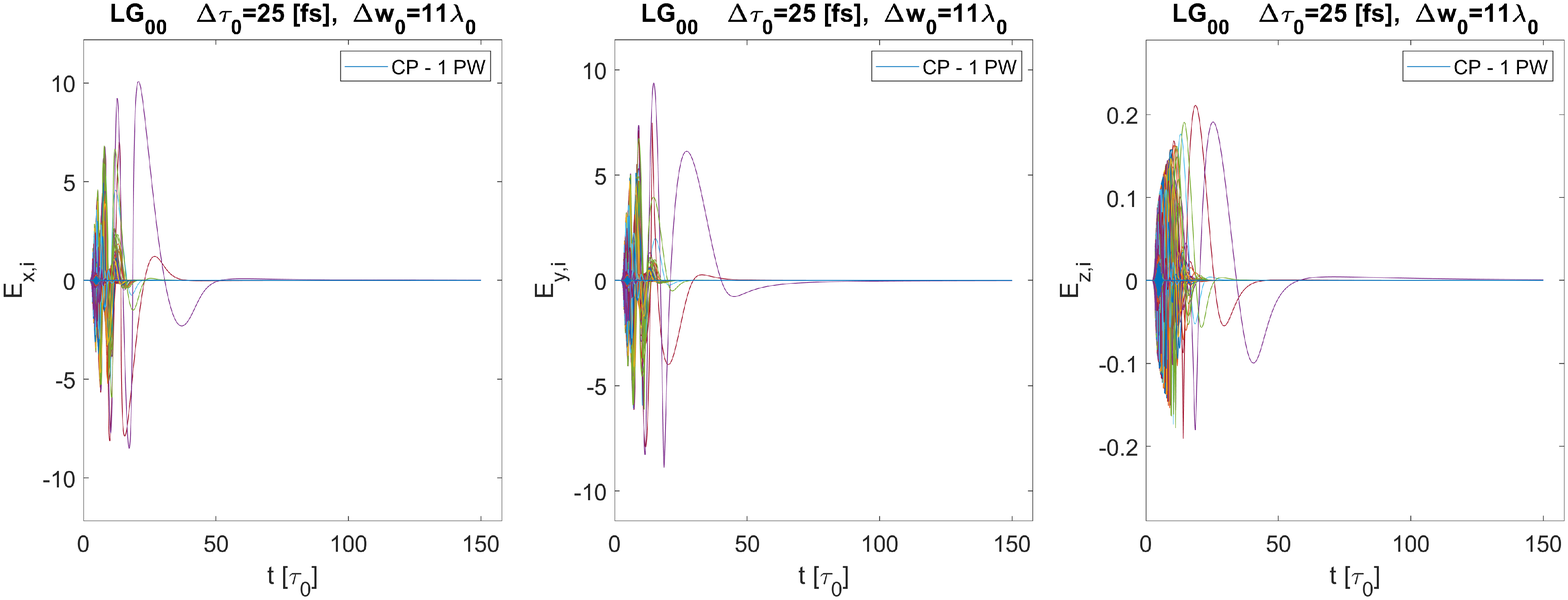} 
\includegraphics[width=17.2cm, height=4.2cm]
{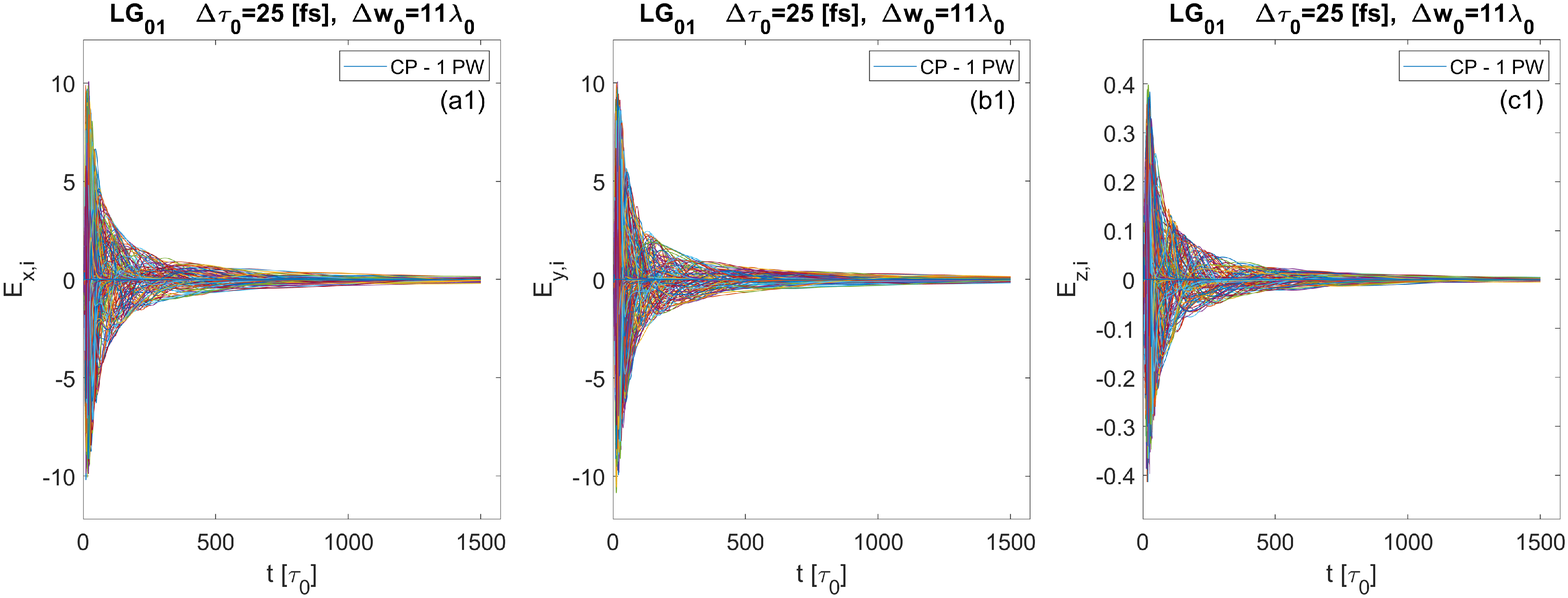} 
\caption{All figures represent CP lasers of $P_0=1$ PW power, with $\Delta \tau_0=25$ fs and
$\Delta w_0=11\protect\lambda_0$.
From left to right: (a0), (b0) and (c0) show the electric fields seen by the electrons $E_{x,i}$, $E_{y,i}$, $E_{z,i}$ in units of $a_0$ as function of time,  
corresponding to the $\text{LG}_{00}$ beam. Similarly, (a1), (b1) and (c1), are for the helical beam $\text{LG}_{01}$.}
\label{fig:Efields_LG0x}
\end{figure*}
\begin{figure*}[hbt!]
\vspace{-0.2cm} 
\includegraphics[width=8.5cm, height=4.4cm]
{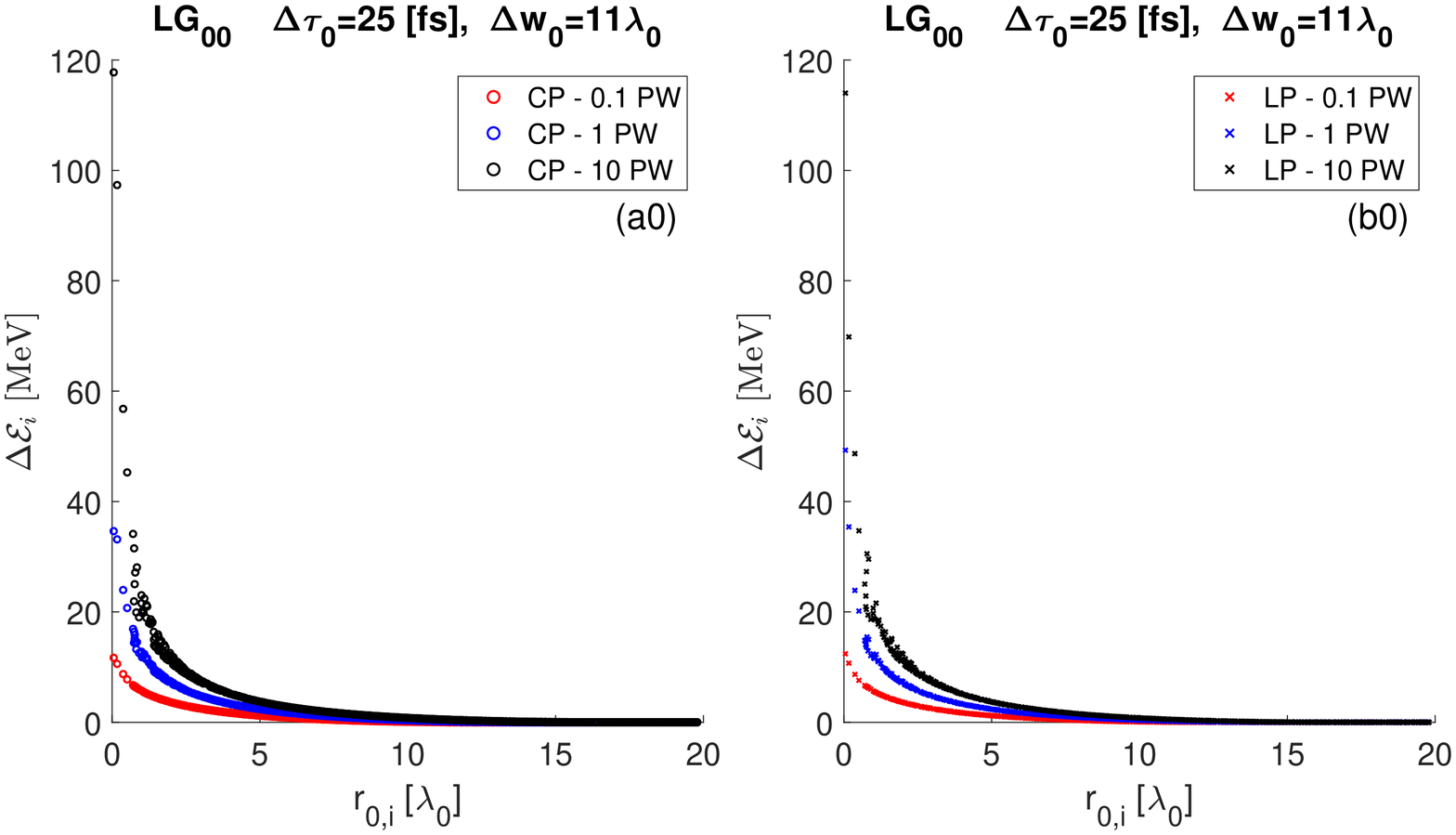} 
\includegraphics[width=8.5cm, height=4.4cm]
{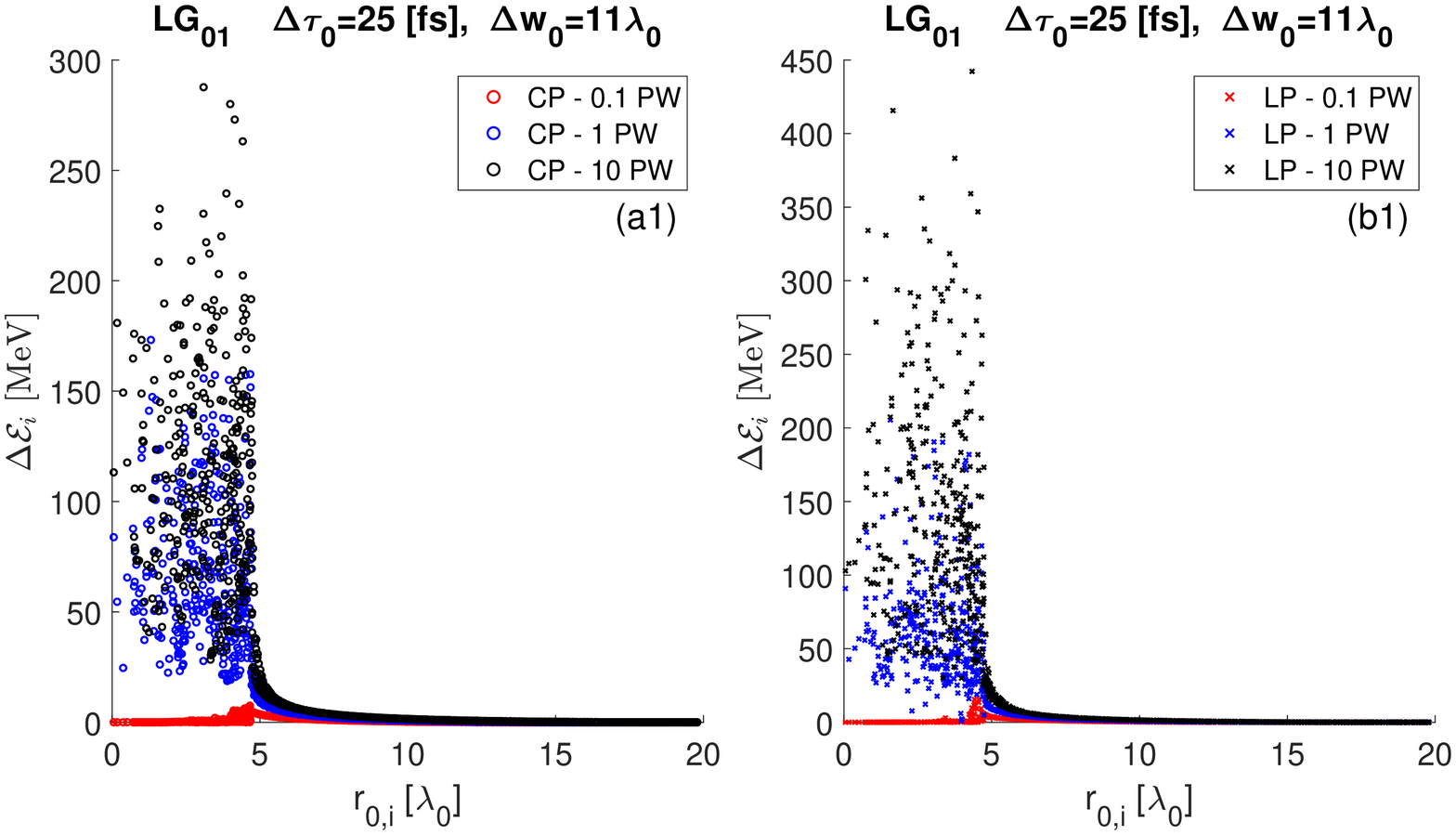} 
\includegraphics[width=8.5cm, height=4.4cm]
{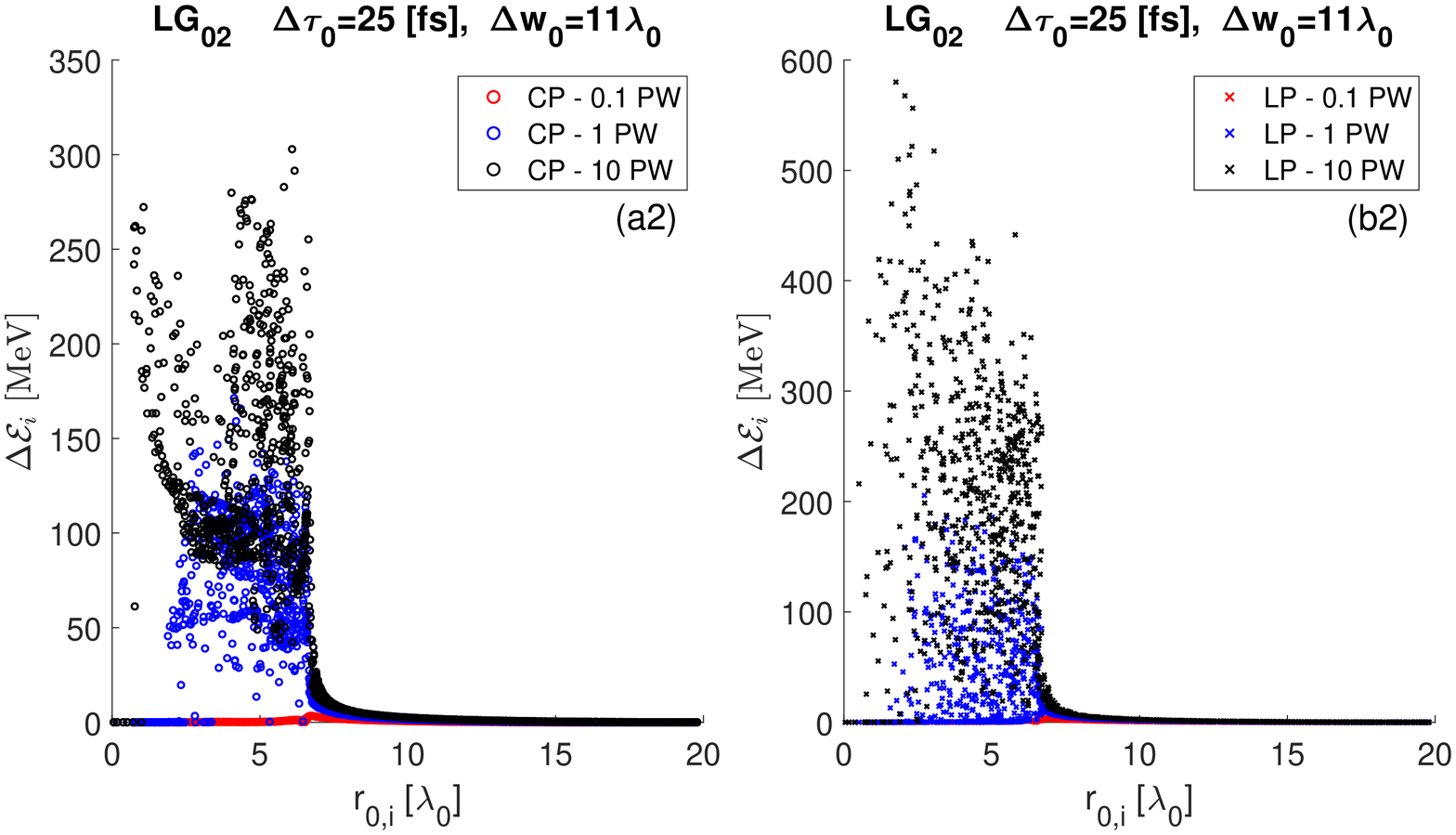} 
\includegraphics[width=8.5cm, height=4.4cm]
{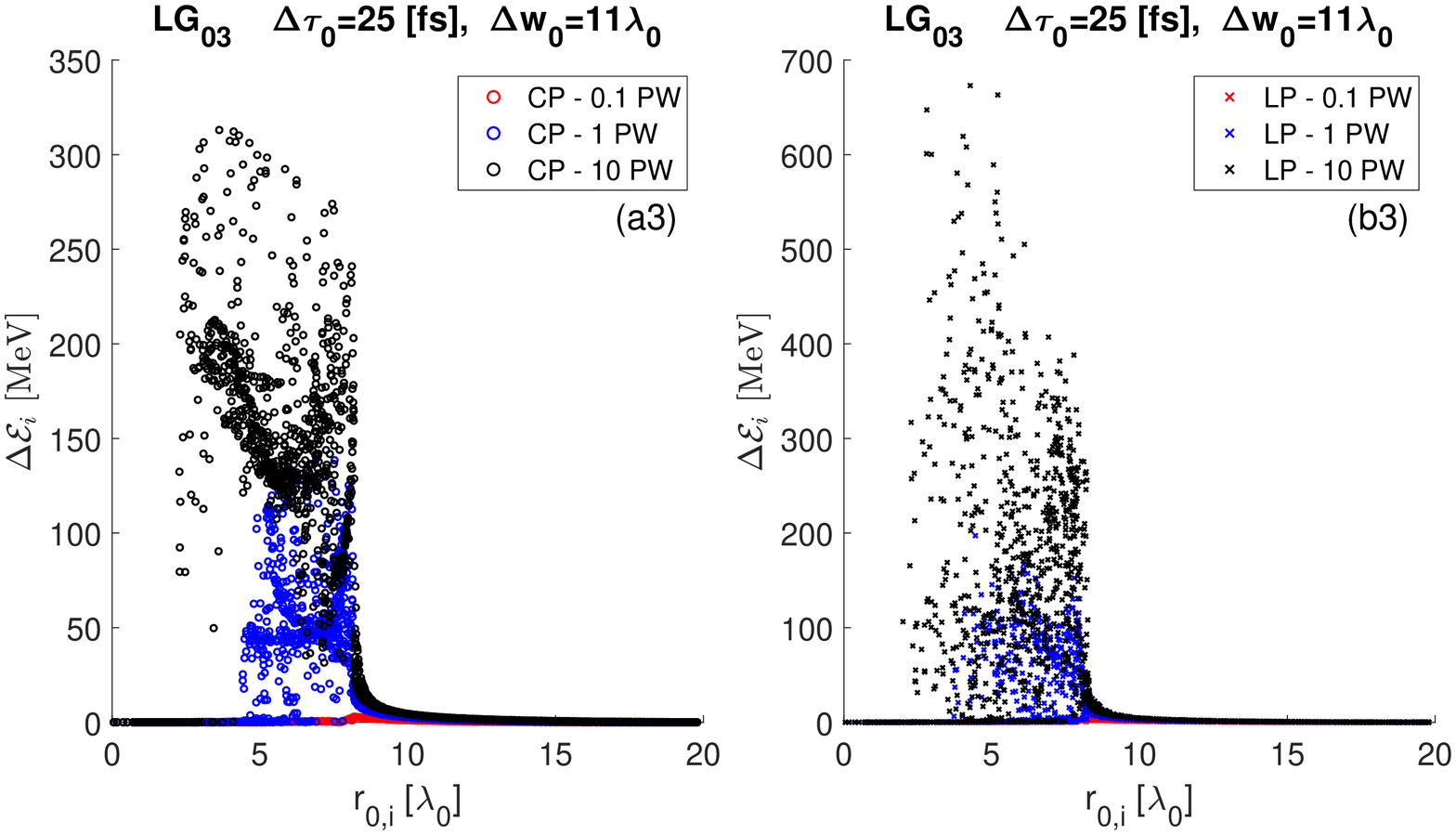}
\includegraphics[width=8.5cm, height=4.4cm]
{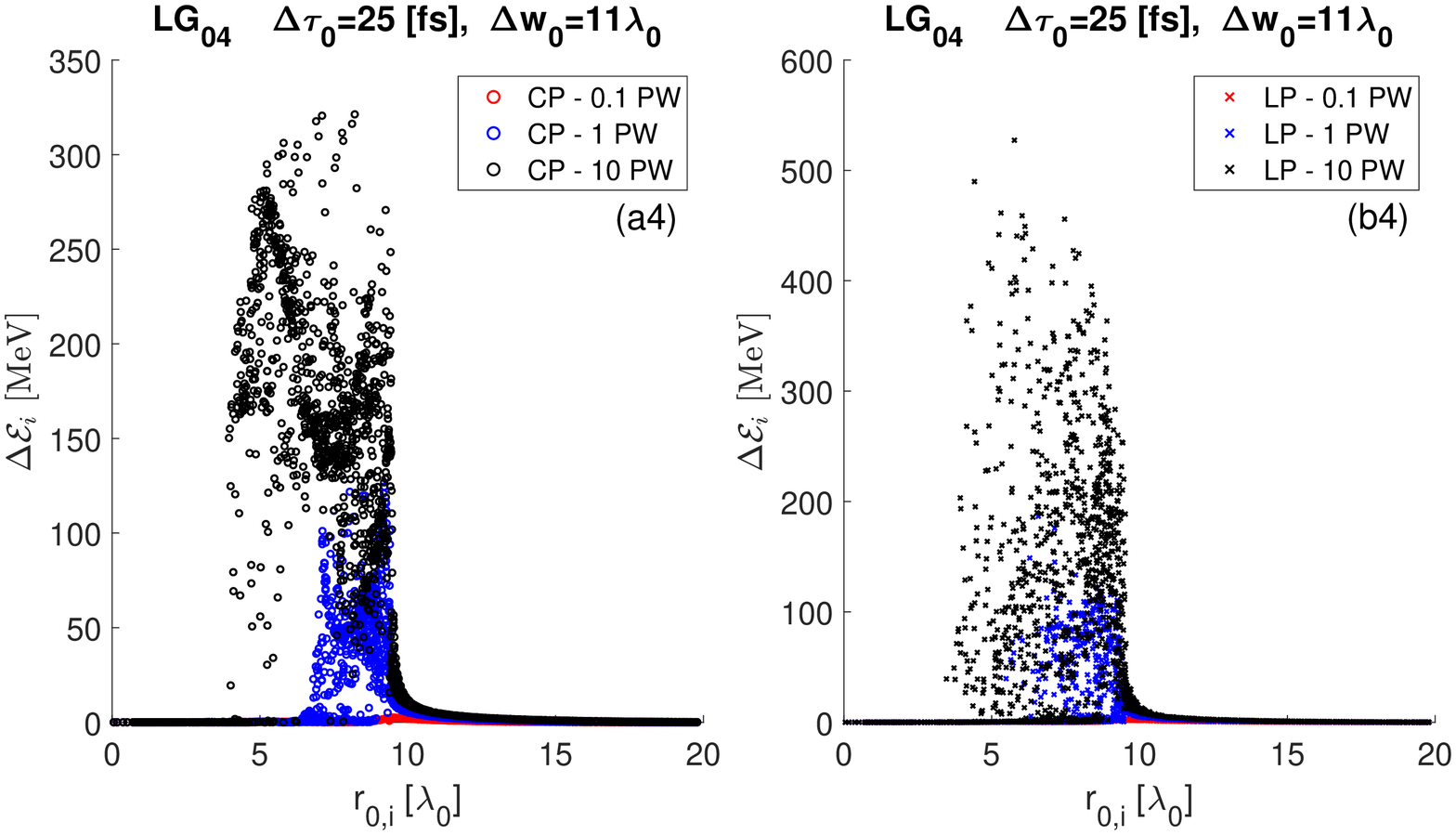}
\includegraphics[width=8.5cm, height=4.4cm]
{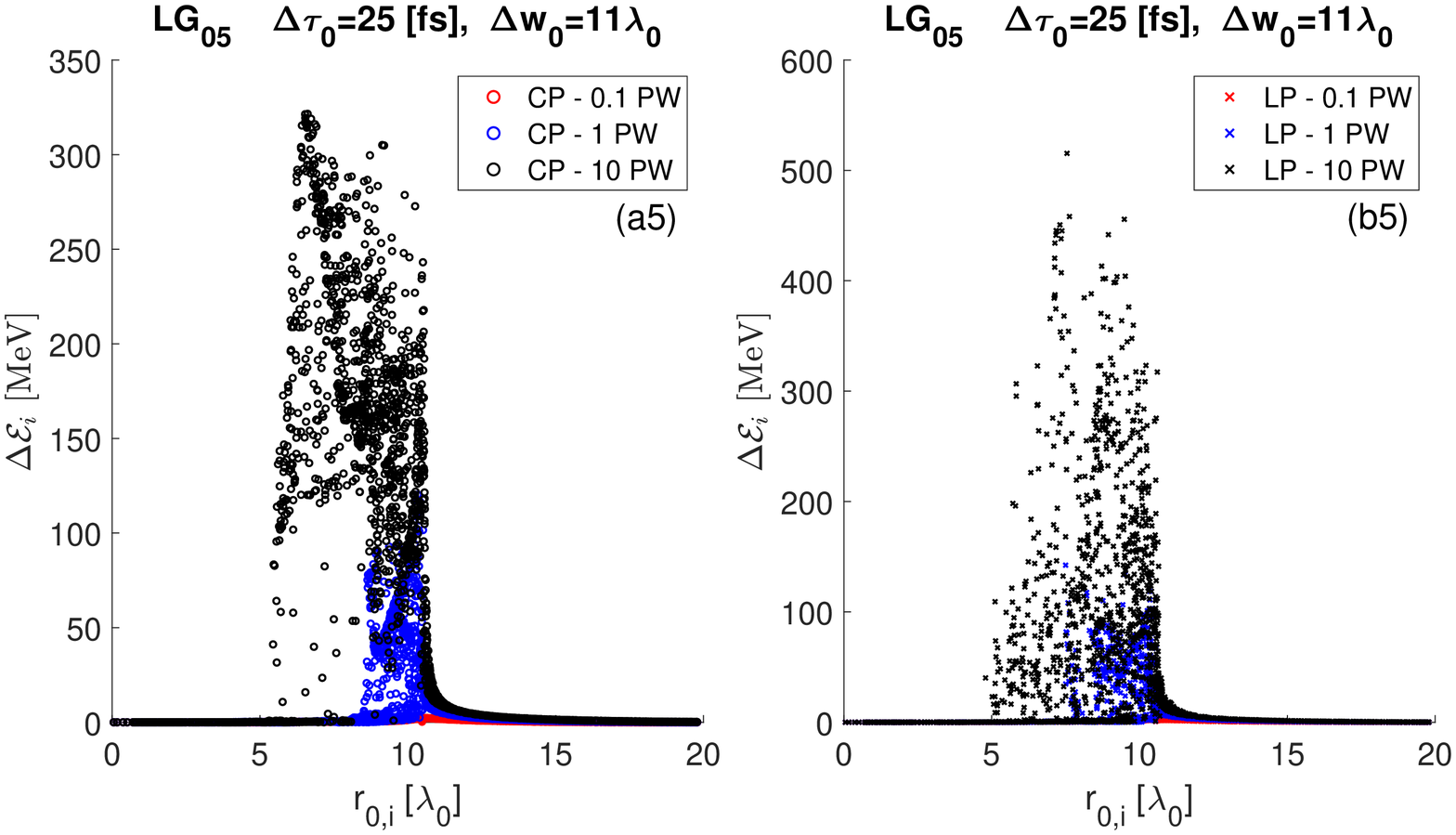}
\caption{All figures show the energy gained 
from laser pulses with $\Delta \tau_0=25$ fs and $\Delta w_0=11\lambda_0$ of $P_0=\left\{0.1, 1, 10\right\}$ PW power with red, blue and black correspondingly, as a function of the same initial radial position of electrons. 
The 1st row from left to right: The 1st and 2nd figures, (a0) and (b0), show the energy gained 
from a Gaussian laser pulse corresponding to CP and LP pulses, which are plotted with "o" and "x" respectively.
The 3rd and 4th figures in the 1st row, (a1) and (b1), show the energy gained as a 
function of the initial radial position in $\text{LG}_{01}$ helical CP and LP beams.
The 2nd row: Similar as before but now the figures, (a2), (b2), and (a3), (b3), correspond to 
$\text{LG}_{02}$ and $\text{LG}_{03}$  helical CP and LP beams.
The 3rd row: Similar as previous rows but now the figures, (a4), (b4), and (a5), (b5), correspond to 
$\text{LG}_{04}$ and $\text{LG}_{05}$ helical CP and LP beams.}
\label{fig:Position_CPLP_LG0x_w011L}
\end{figure*}
\begin{figure*}[hbt!]
\vspace{-0.2cm} 
\includegraphics[width=8.5cm, height=4.2cm]
{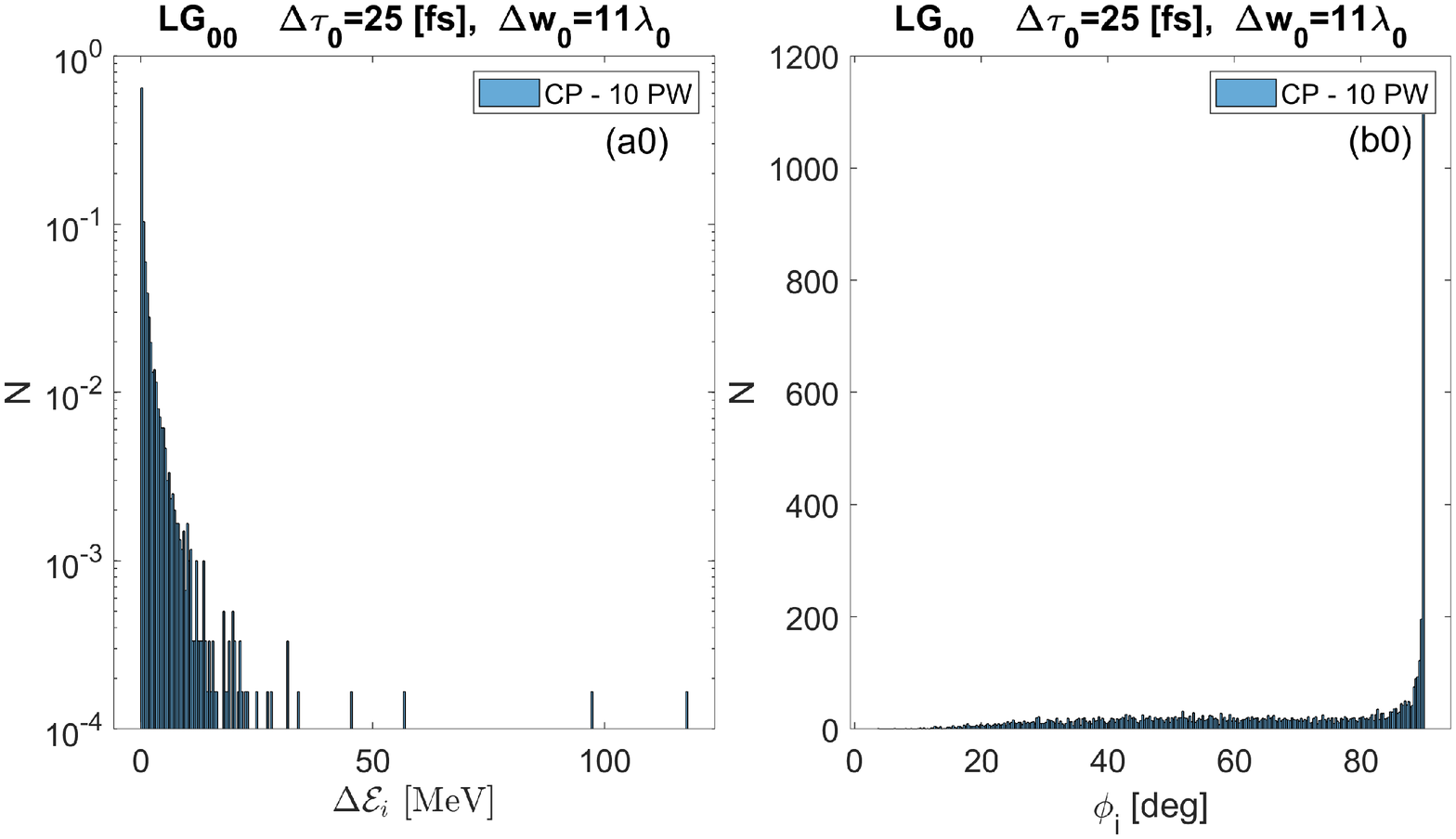} 
\includegraphics[width=8.5cm, height=4.2cm]
{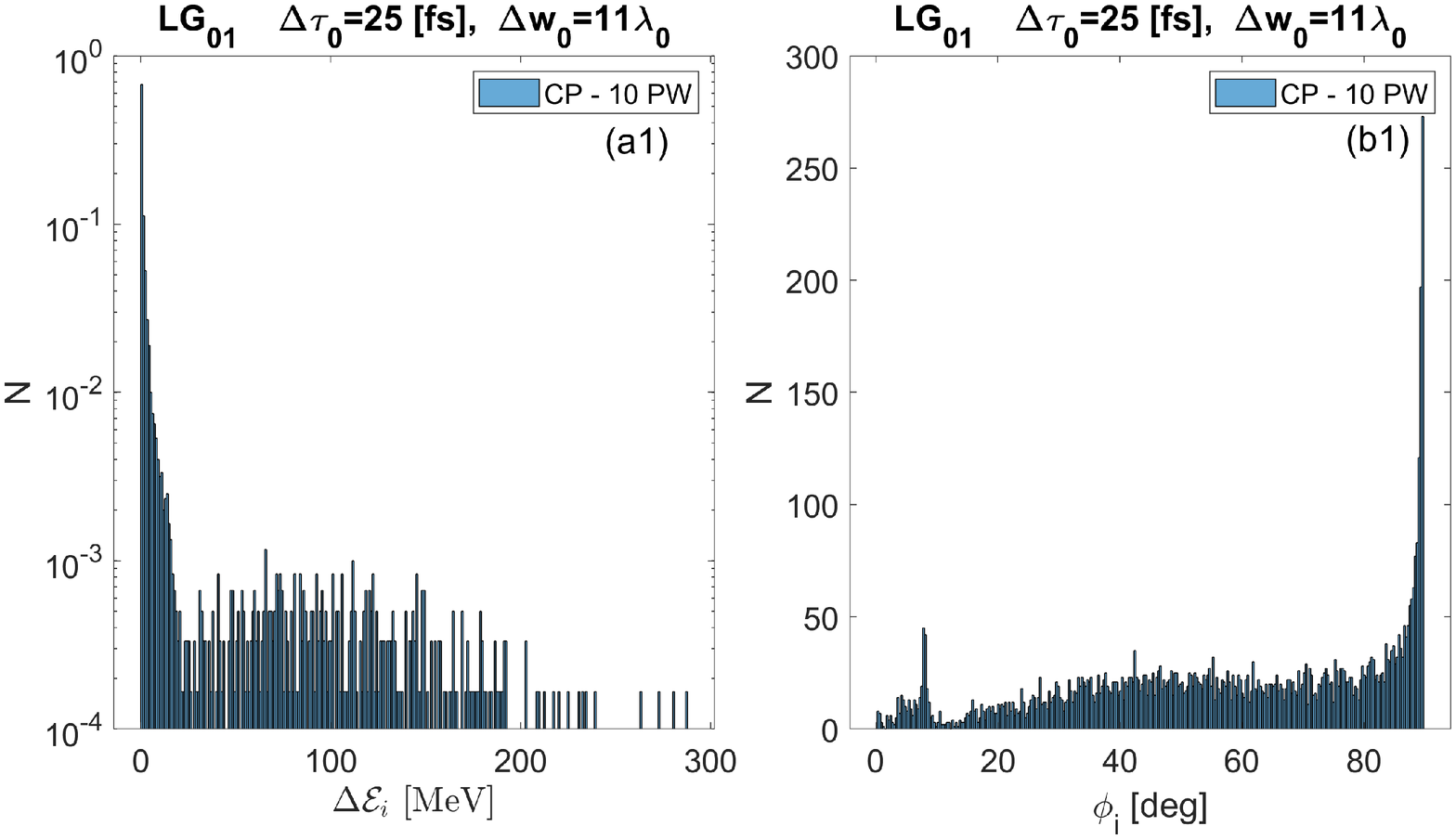} 
\includegraphics[width=8.5cm, height=4.2cm]
{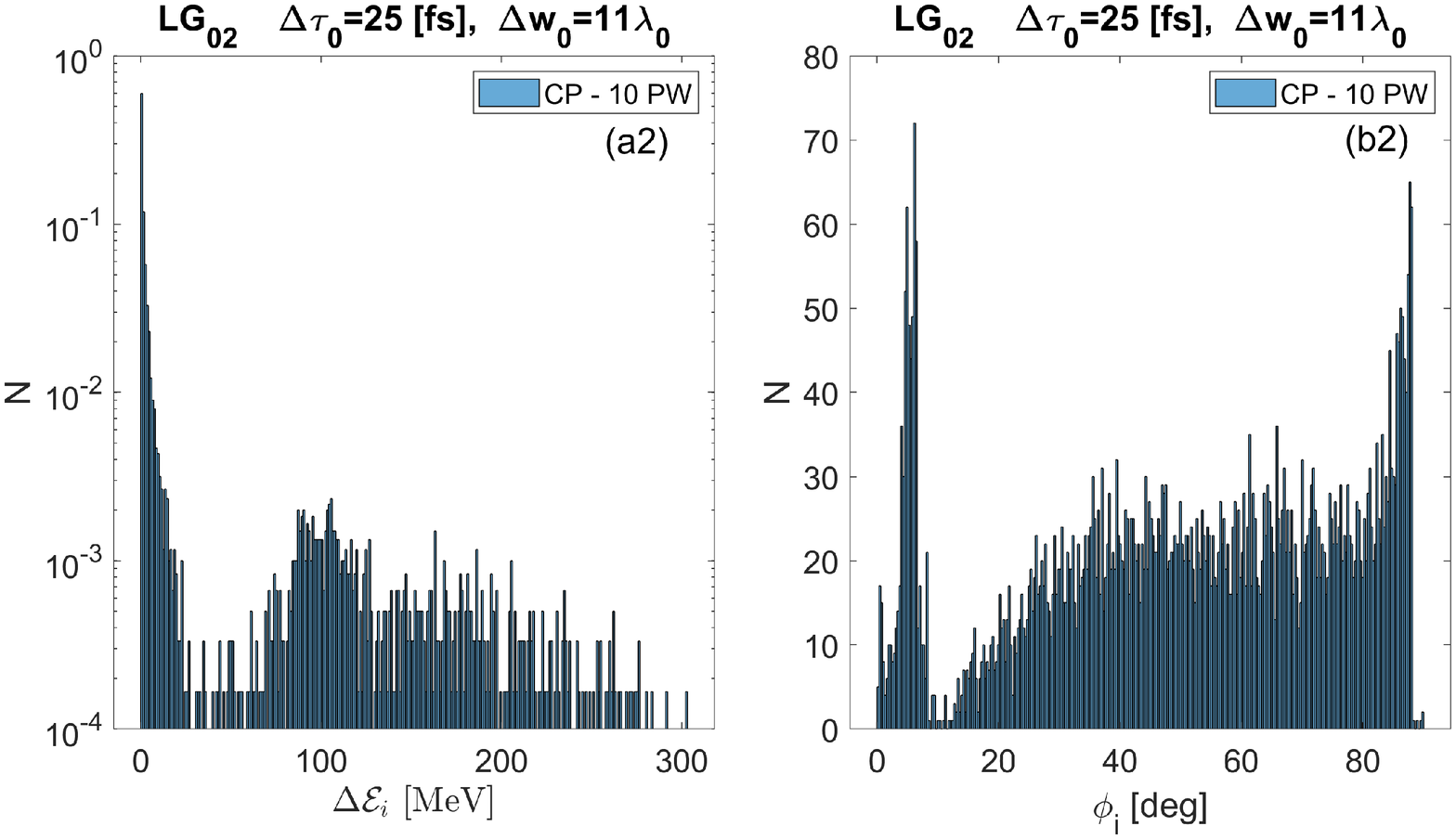} 
\includegraphics[width=8.5cm, height=4.2cm]
{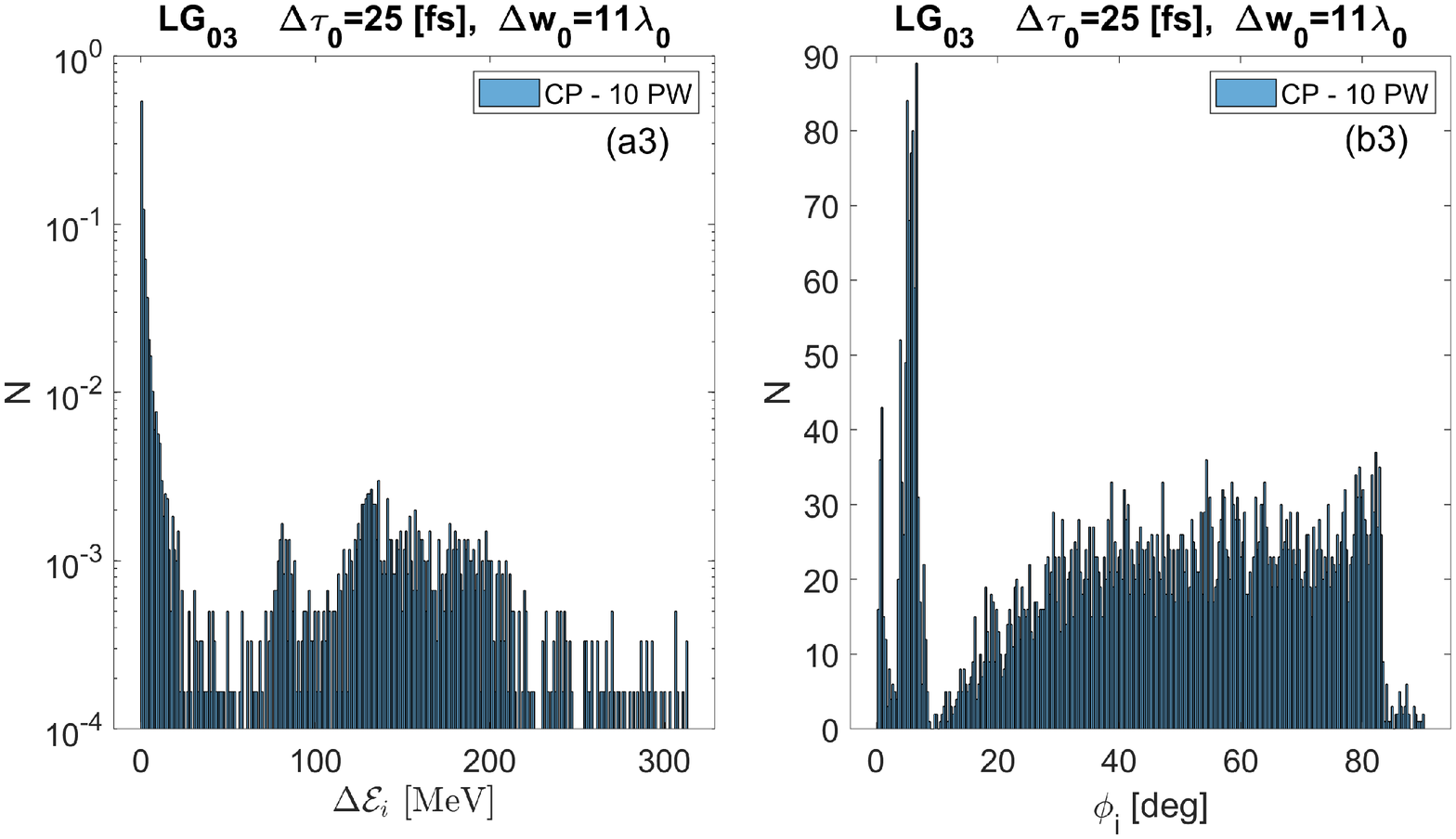}
\includegraphics[width=8.5cm, height=4.2cm]
{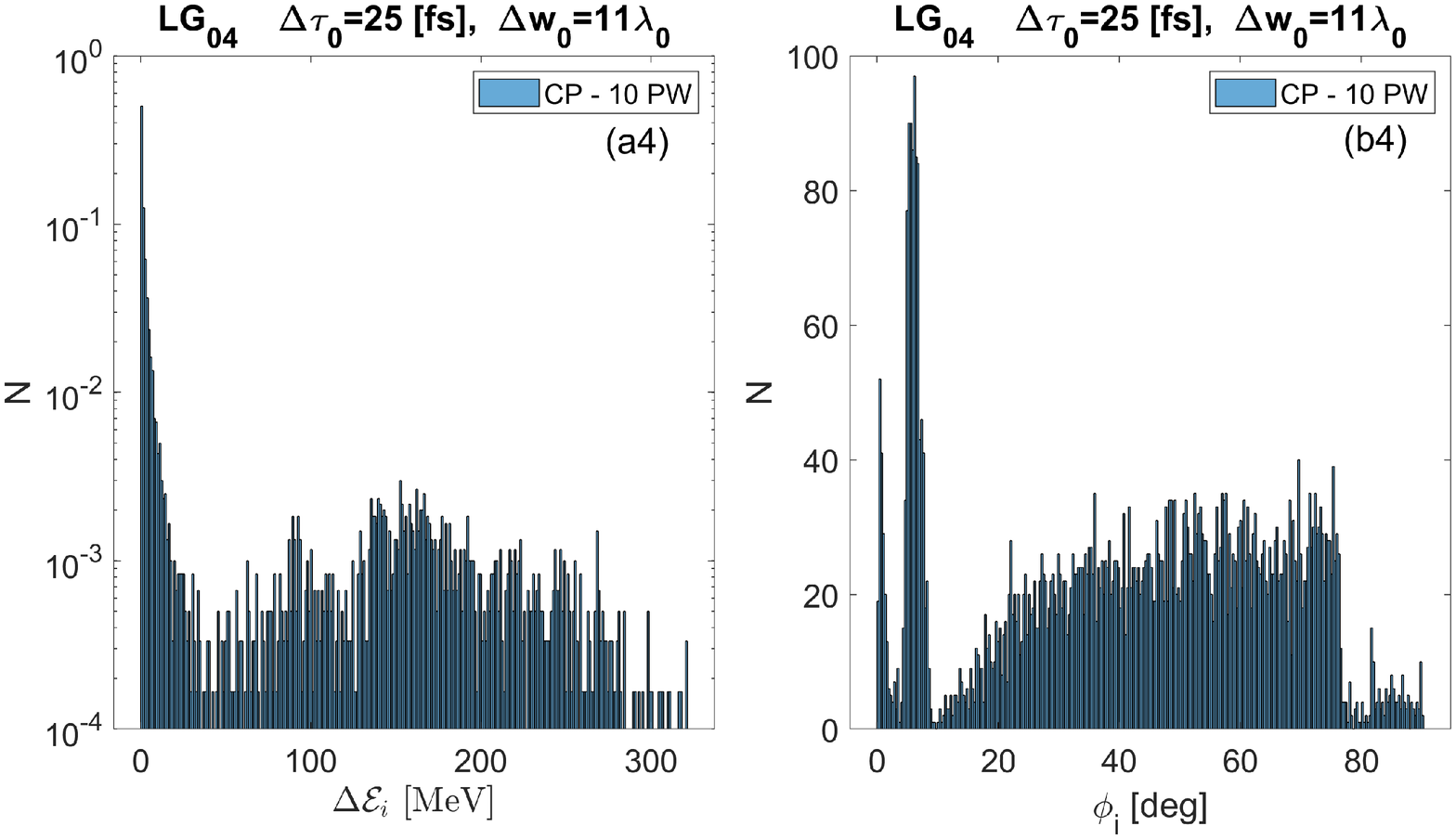}
\includegraphics[width=8.5cm, height=4.2cm]
{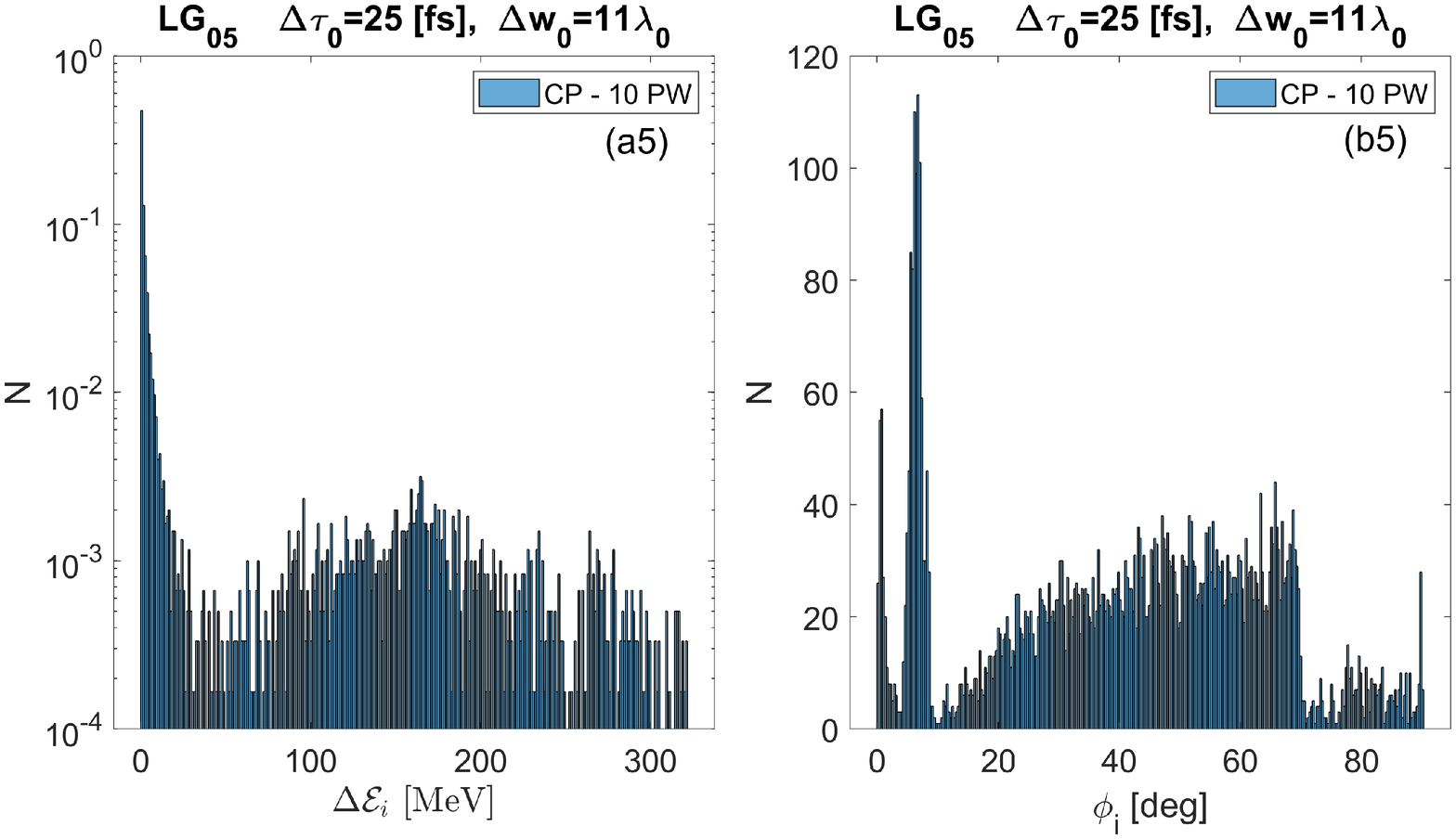}
\caption{The figures show the logarithmic scale histogram of net energy gained $\Delta \mathcal{E}_i$ and the 
polar angle histogram $\phi_i$ of electrons after interaction with a circularly polarized laser pulse
with $\Delta \tau_0=25$ fs, $\Delta w_0=11\lambda_0$ beam waist and $P_0 = 10$ PW power.
The 1st row from left to right: The 1st and 2nd figures, (a0) and (b0), show the energy gained 
form a Gaussian pulse and the corresponding angular distribution of electrons.
Similarly as before, the 3rd and 4th figures in the 1st row, (a1) and (b1), show the energy gain 
and angular distribution in case of the $\text{LG}_{01}$ helical beam.
The 2nd row: Similar as the 1st row, but now figures (a2), (b2), and (a3), (b3), are for the helical beams 
$\text{LG}_{02}$ and $\text{LG}_{03}$. 
The 3rd row: Similar as previous rows, but now figures (a4), (b4), and (a5), (b5), are for the helical beams 
$\text{LG}_{04}$ and $\text{LG}_{05}$.}
\label{fig:Histogram_CPLP_LG0x_w011L}
\end{figure*}
\begin{figure*}[hbt!]
\vspace{-0.2cm} 
\includegraphics[width=8.5cm, height=4.4cm]
{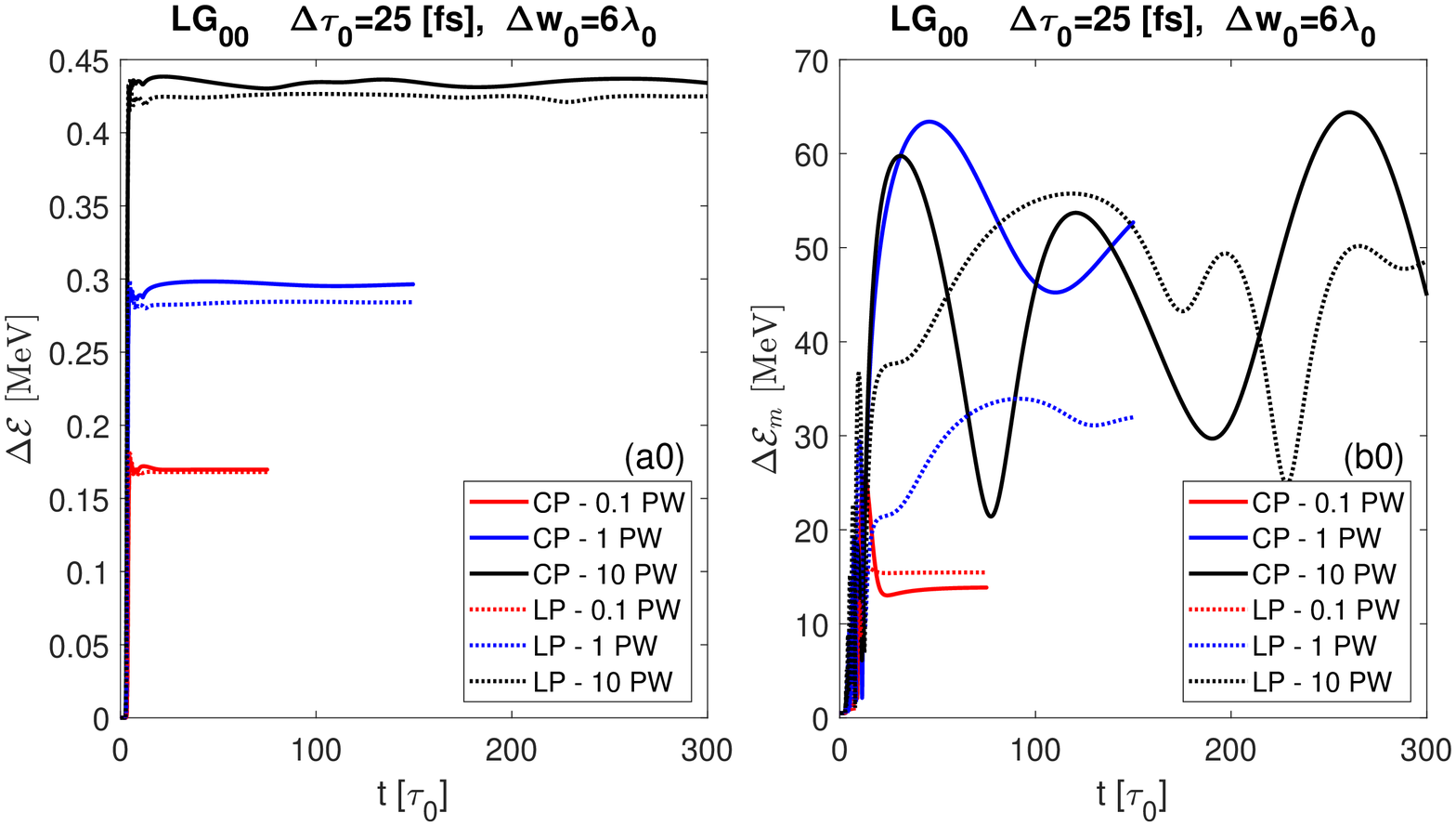} 
\includegraphics[width=8.5cm, height=4.4cm]
{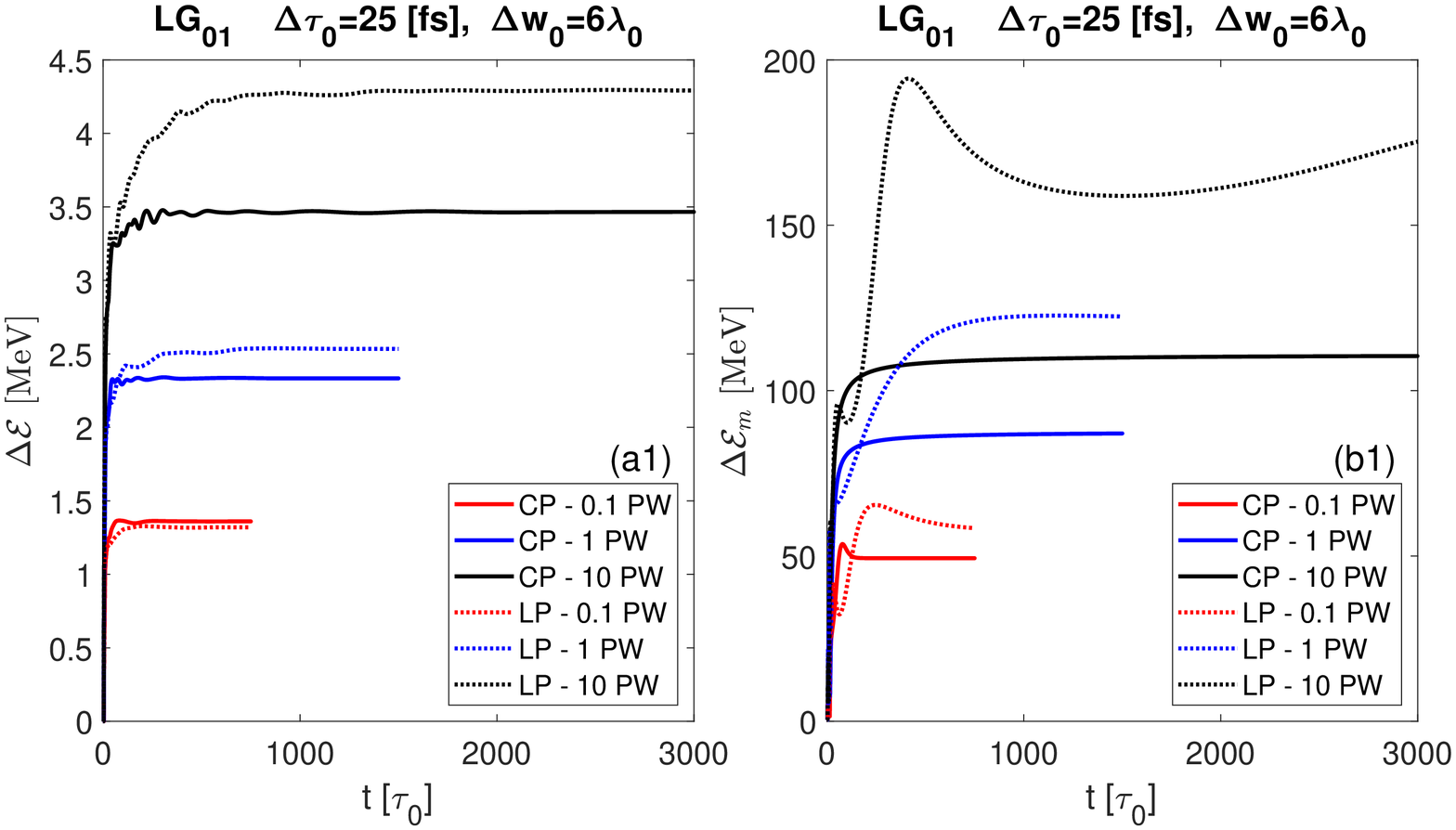} 
\includegraphics[width=8.5cm, height=4.4cm]
{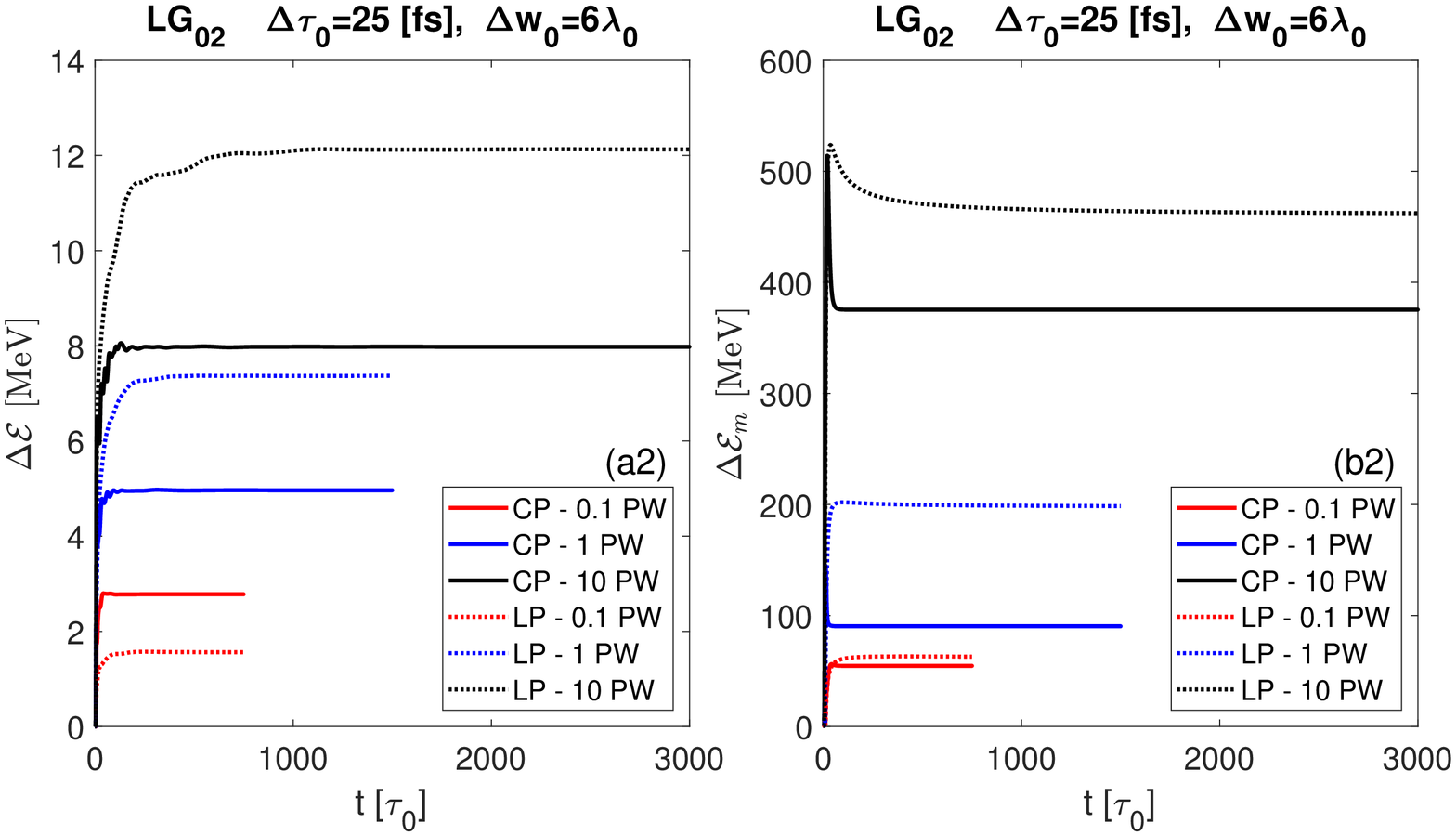} 
\includegraphics[width=8.5cm, height=4.4cm]
{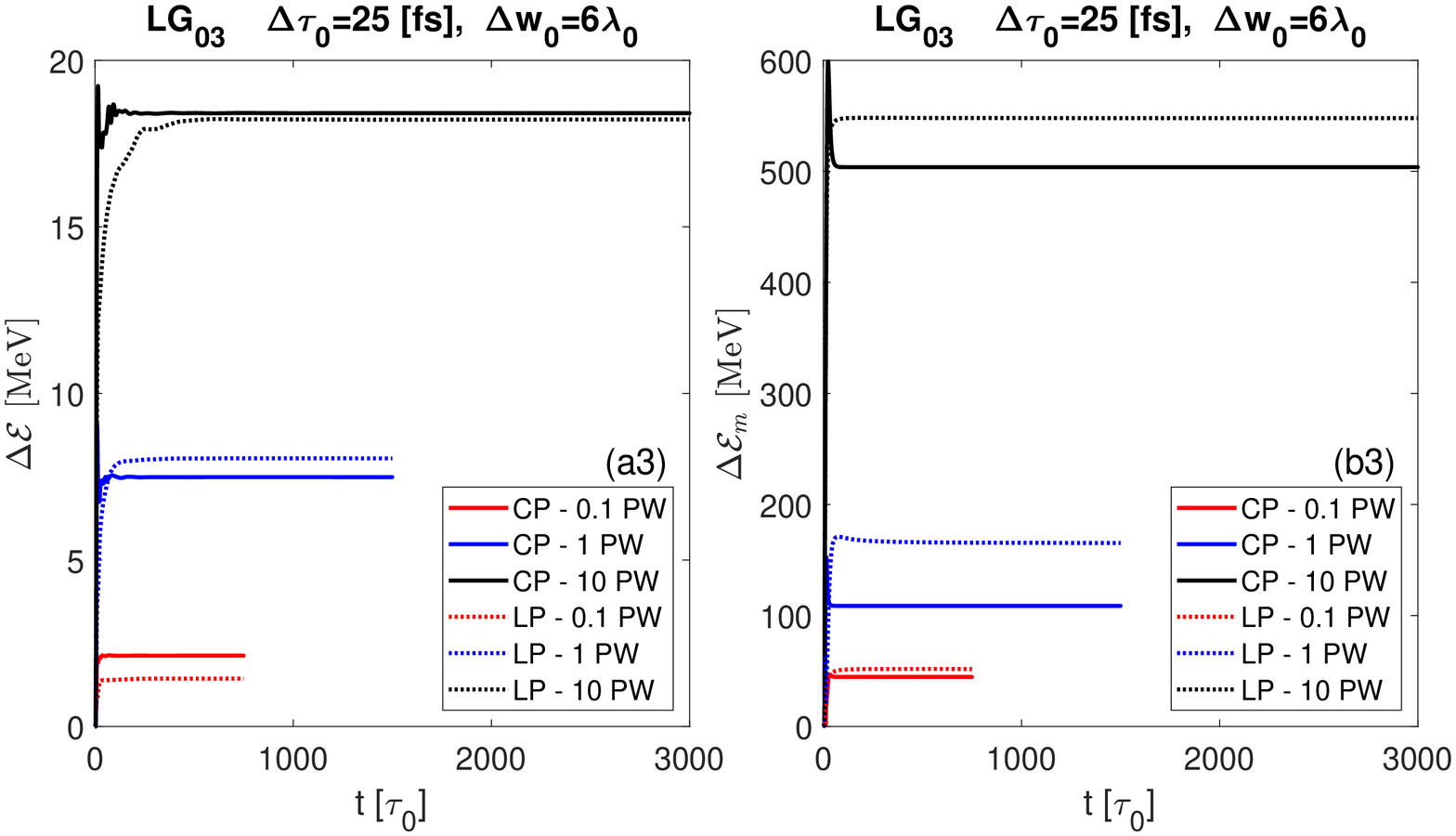}
\includegraphics[width=8.5cm, height=4.4cm]
{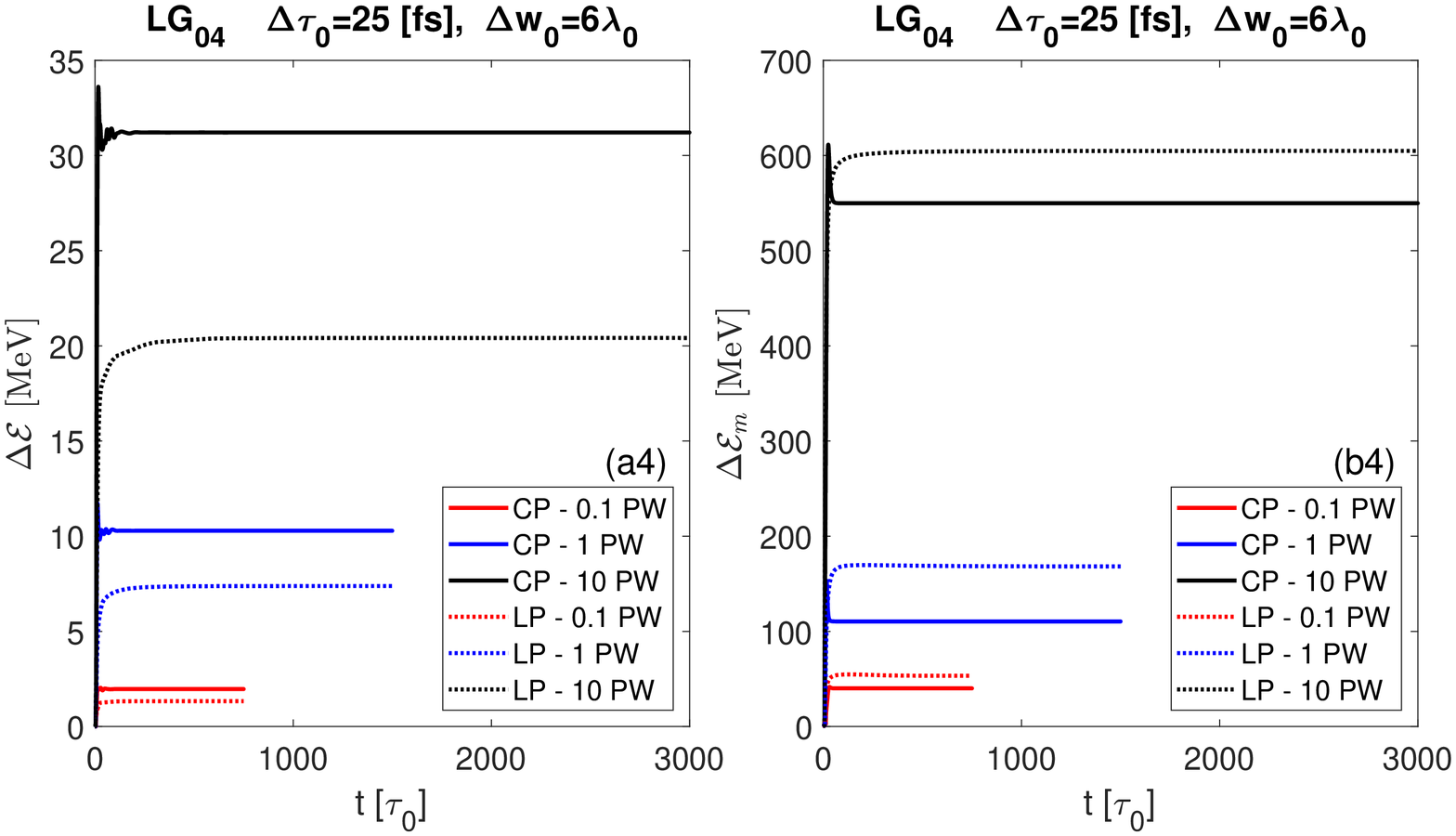}
\includegraphics[width=8.5cm, height=4.4cm]
{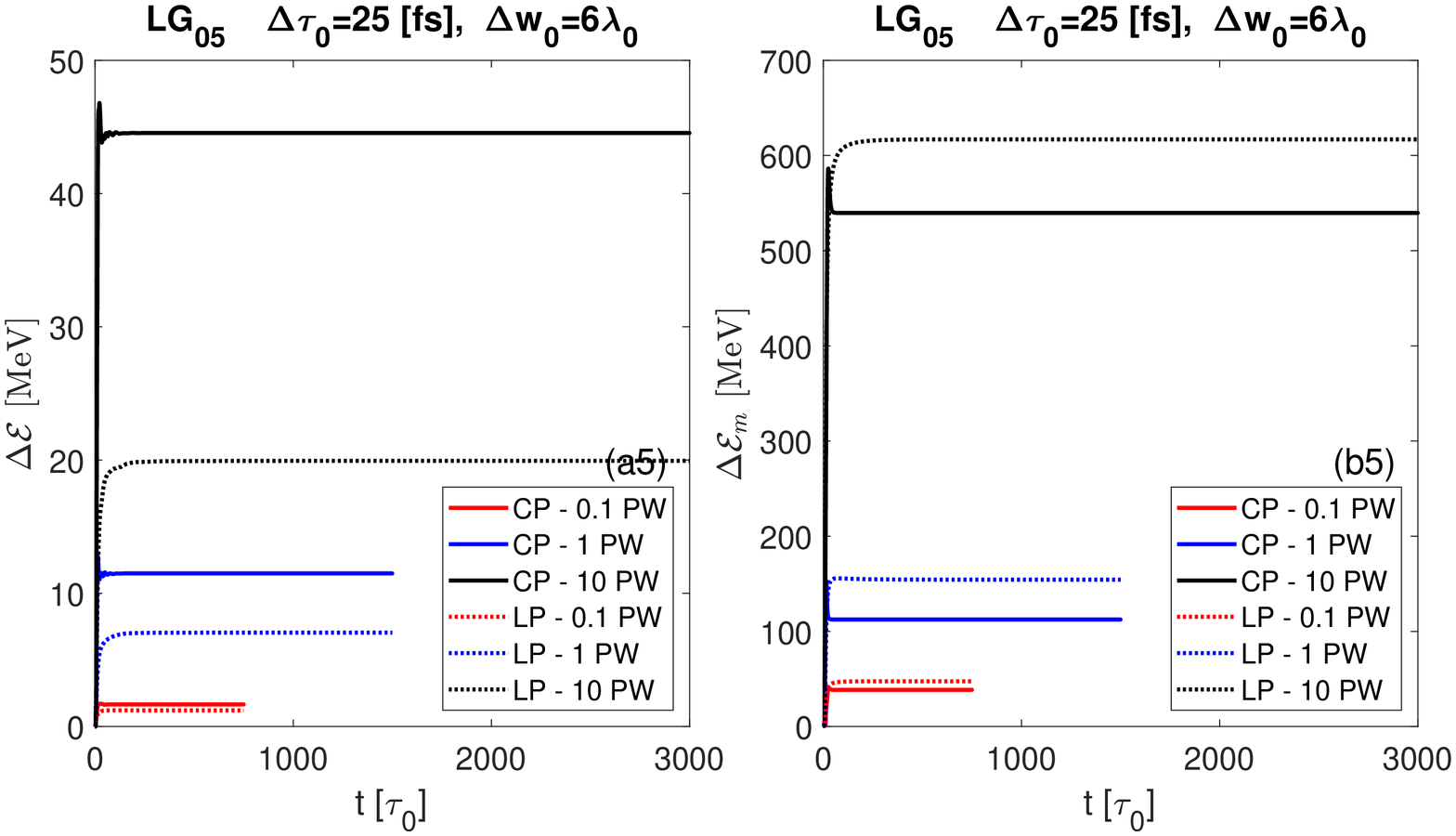}
\caption{All figures correspond to lasers with $\Delta \tau_0=25$ fs pulse duration and $\Delta w_0=6 \lambda_0$ beam waist radius. 
The full and dotted lines correspond to a CP and LP laser  
of $P_0=\left\{0.1, 1, 10\right\}$ PW power, with red, blue and black, respectively.
The 1st row from left to right: The 1st figure (a0) shows the time evolution of the mean net energy gain 
of electrons, while the 2nd figure (b0) shows the time evolution of the highest energy 
electron, $\Delta \mathcal{E}_m$, both corresponding to Gaussian pulses of different powers and polarization.
Similarly as before: the 3rd and 4th figures, (a1) and (b1), 
show the energy gains corresponding to the $\text{LG}_{01}$ helical beam.
The 2nd row: Similar as before, but now figures (a2), (b2), and (a3), (b3), correspond to $\text{LG}_{02}$ 
and $\text{LG}_{03}$ helical beams.
The 3rd row: figures (a4), (b4), and (a5), (b5), correspond to $\text{LG}_{04}$ 
and $\text{LG}_{05}$ helical beams.}
\label{fig:Energy_CPLP_LG0x_w06L}
\end{figure*}
\begin{figure*}[hbt!]
\vspace{-0.2cm} 
\includegraphics[width=8.5cm, height=4.4cm]
{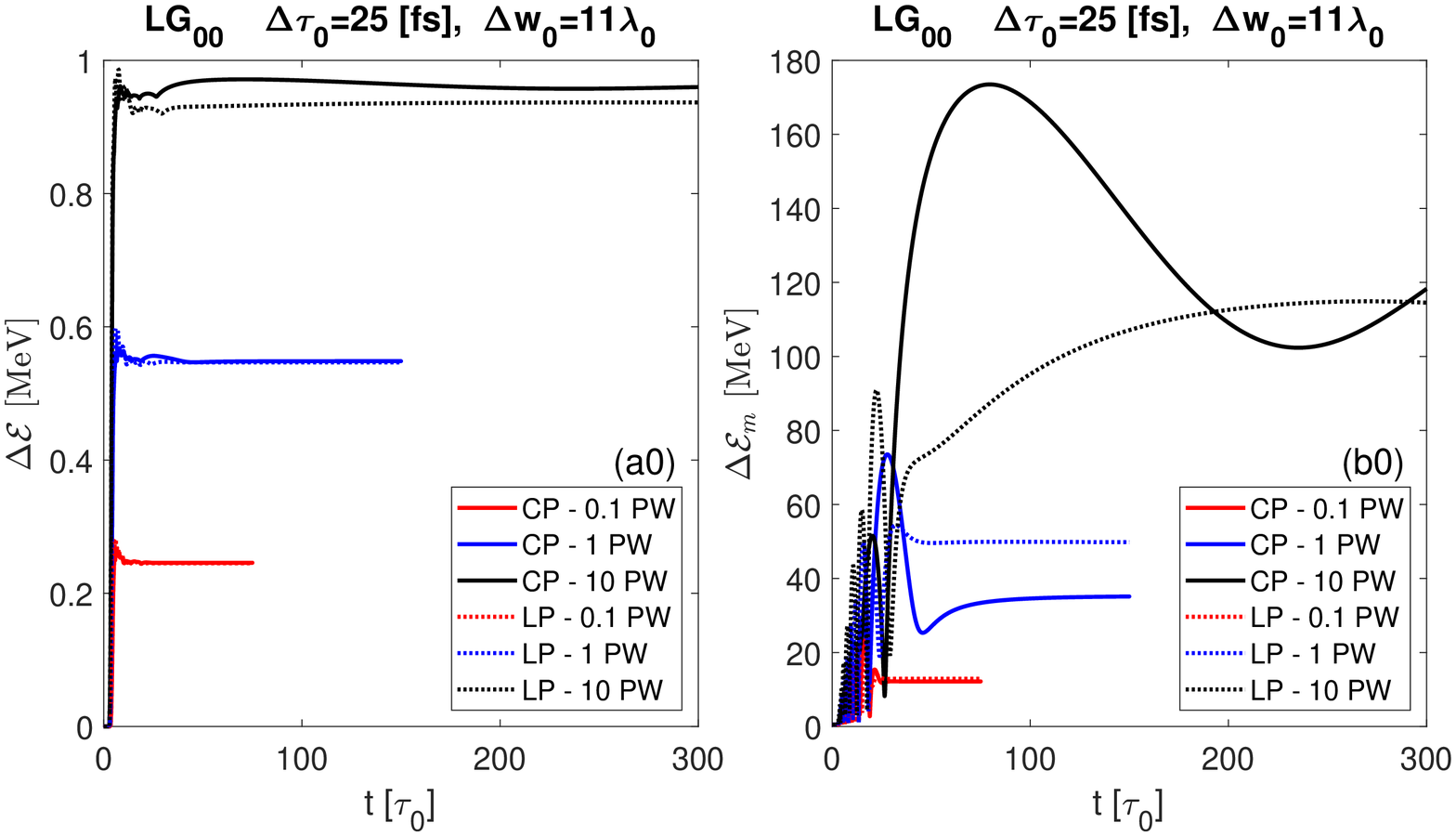} 
\includegraphics[width=8.5cm, height=4.4cm]
{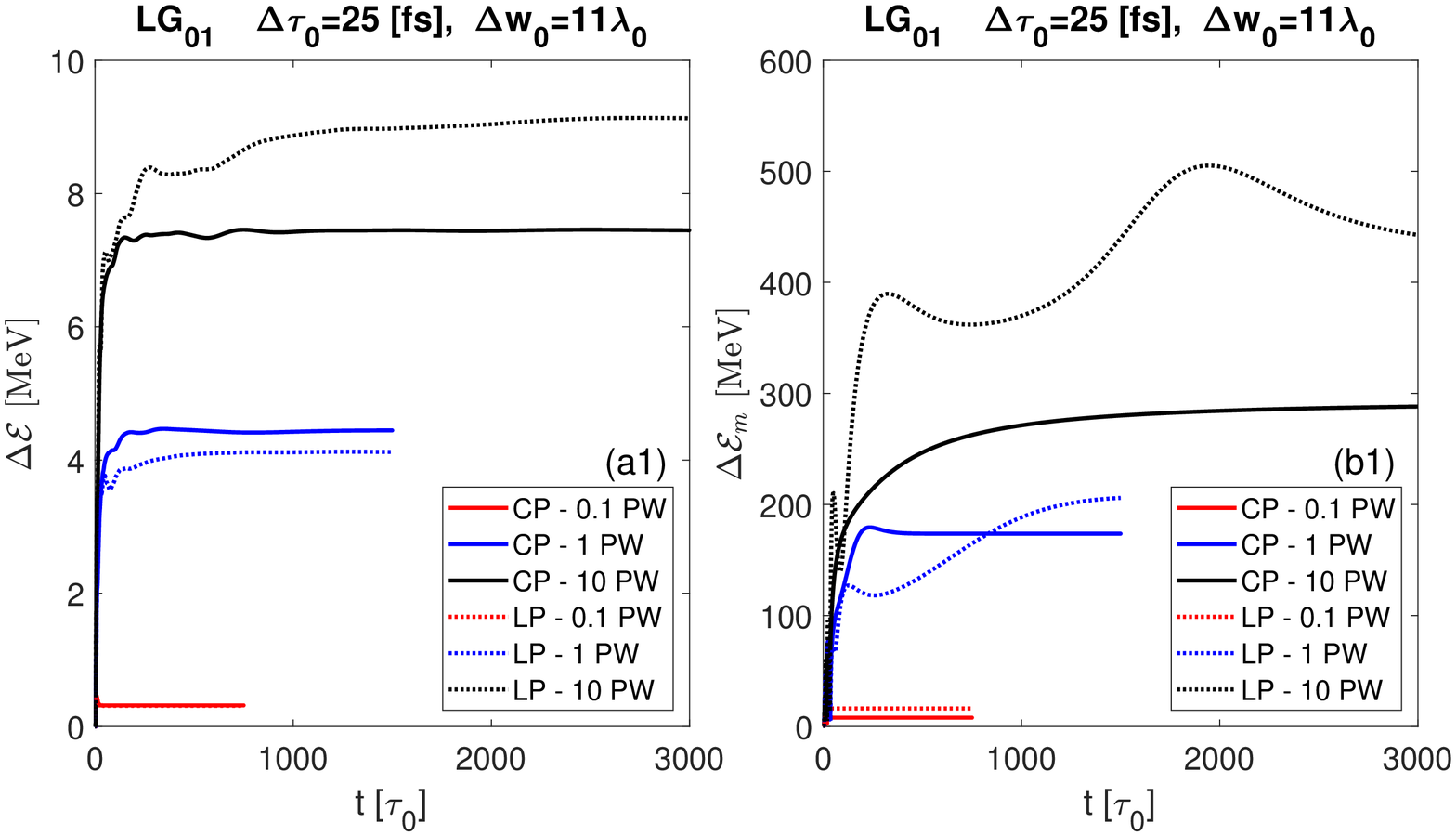} 
\includegraphics[width=8.5cm, height=4.4cm]
{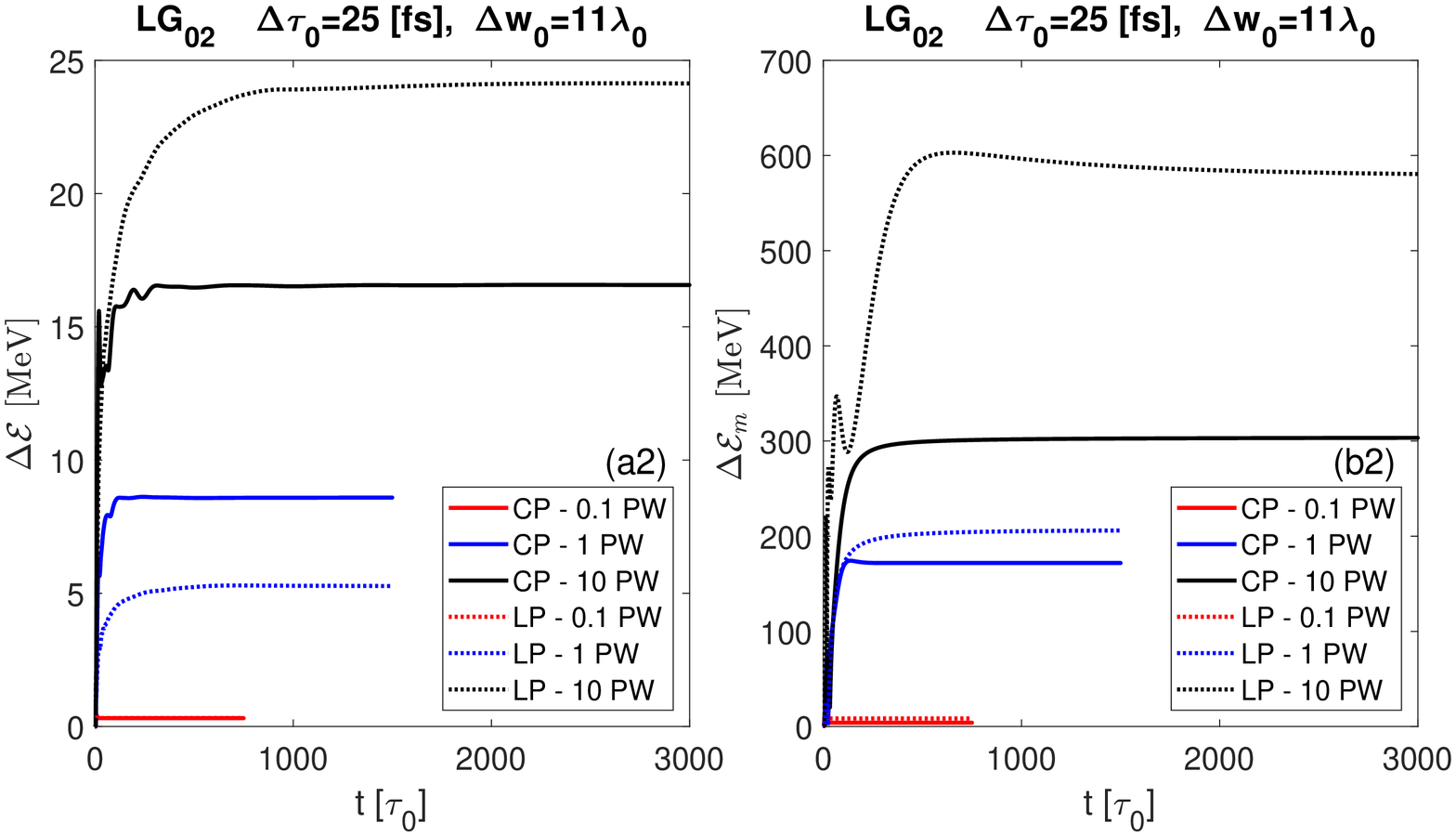} 
\includegraphics[width=8.5cm, height=4.4cm]
{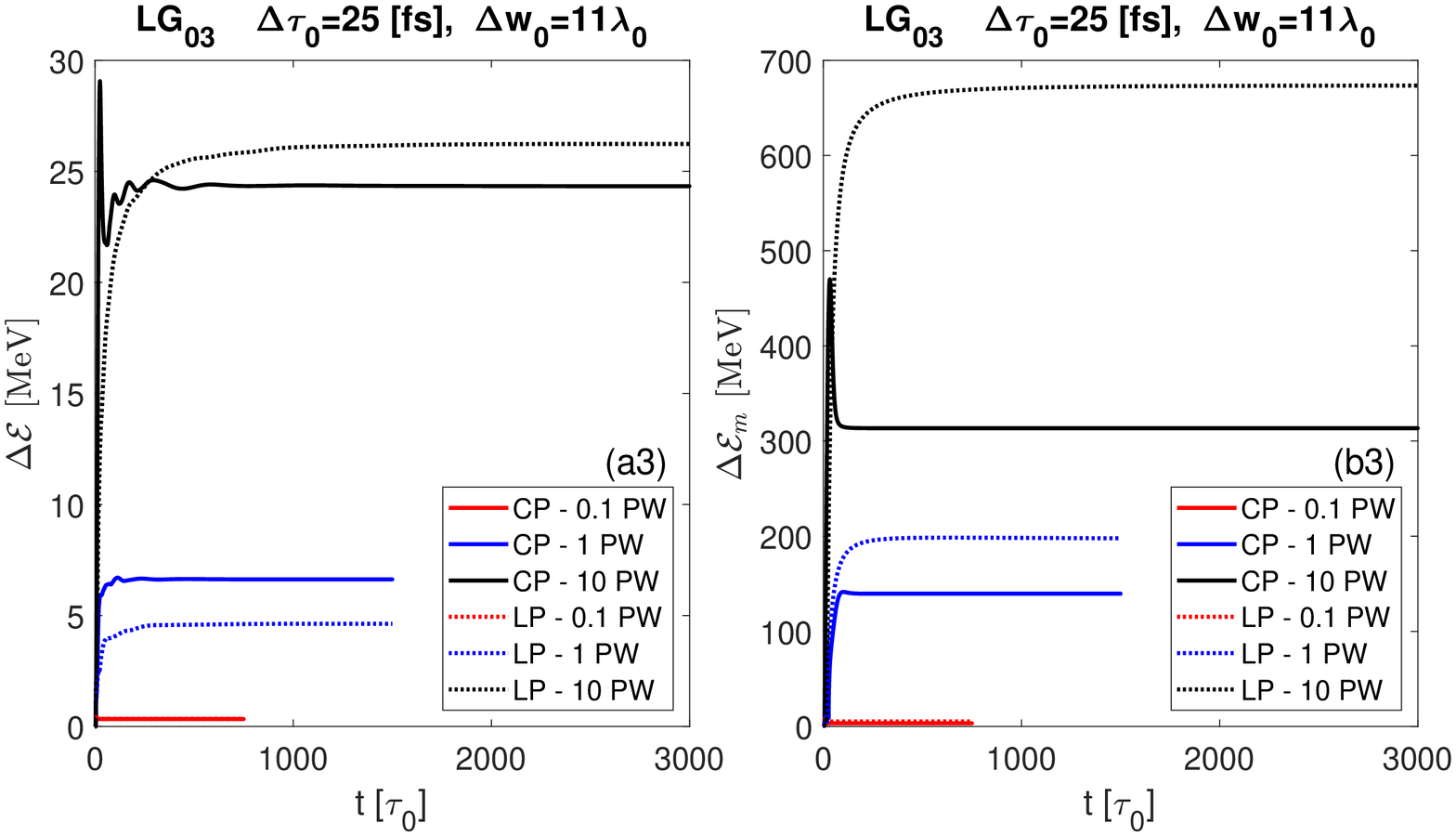}
\includegraphics[width=8.5cm, height=4.4cm]
{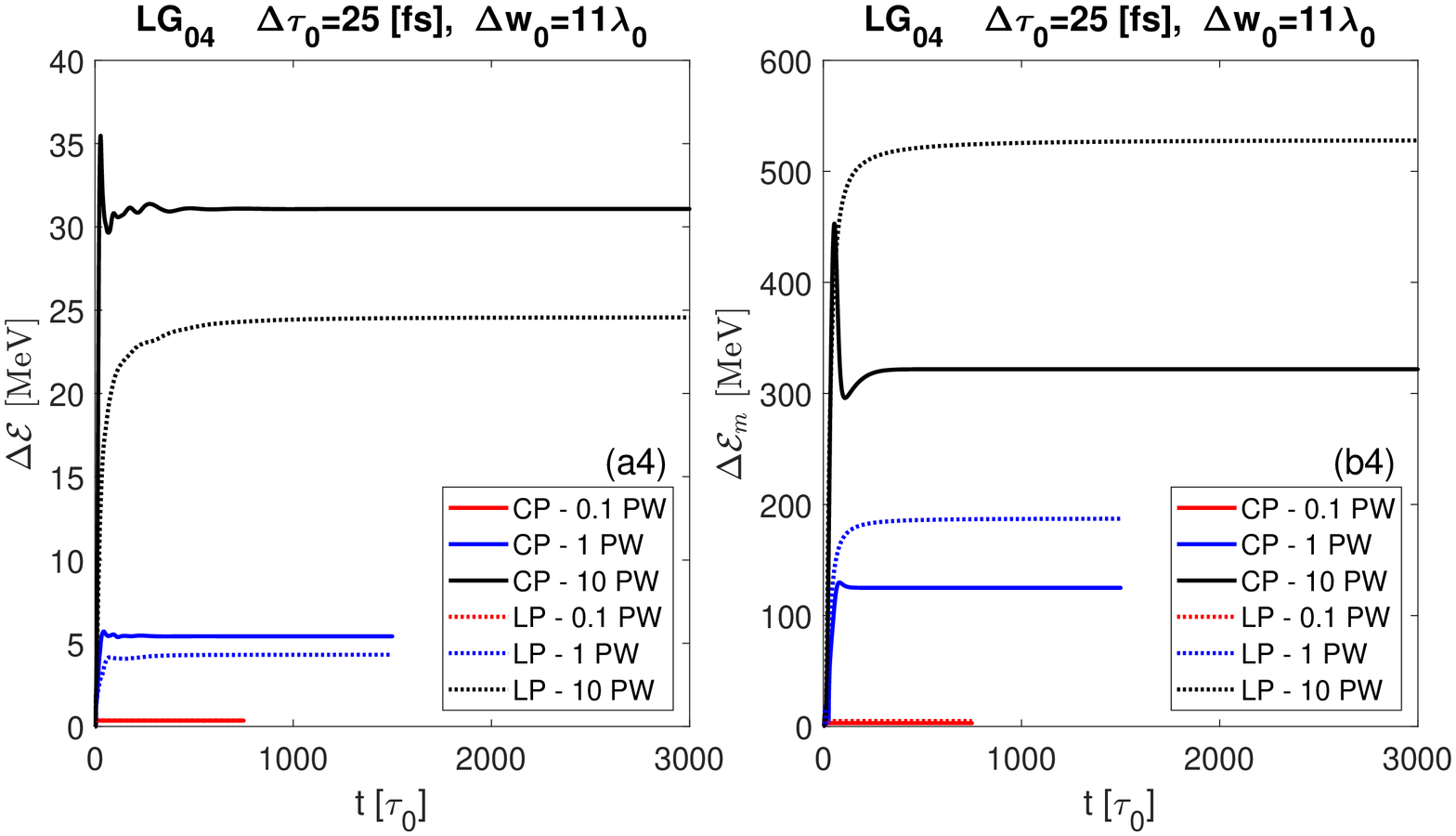}
\includegraphics[width=8.5cm, height=4.4cm]
{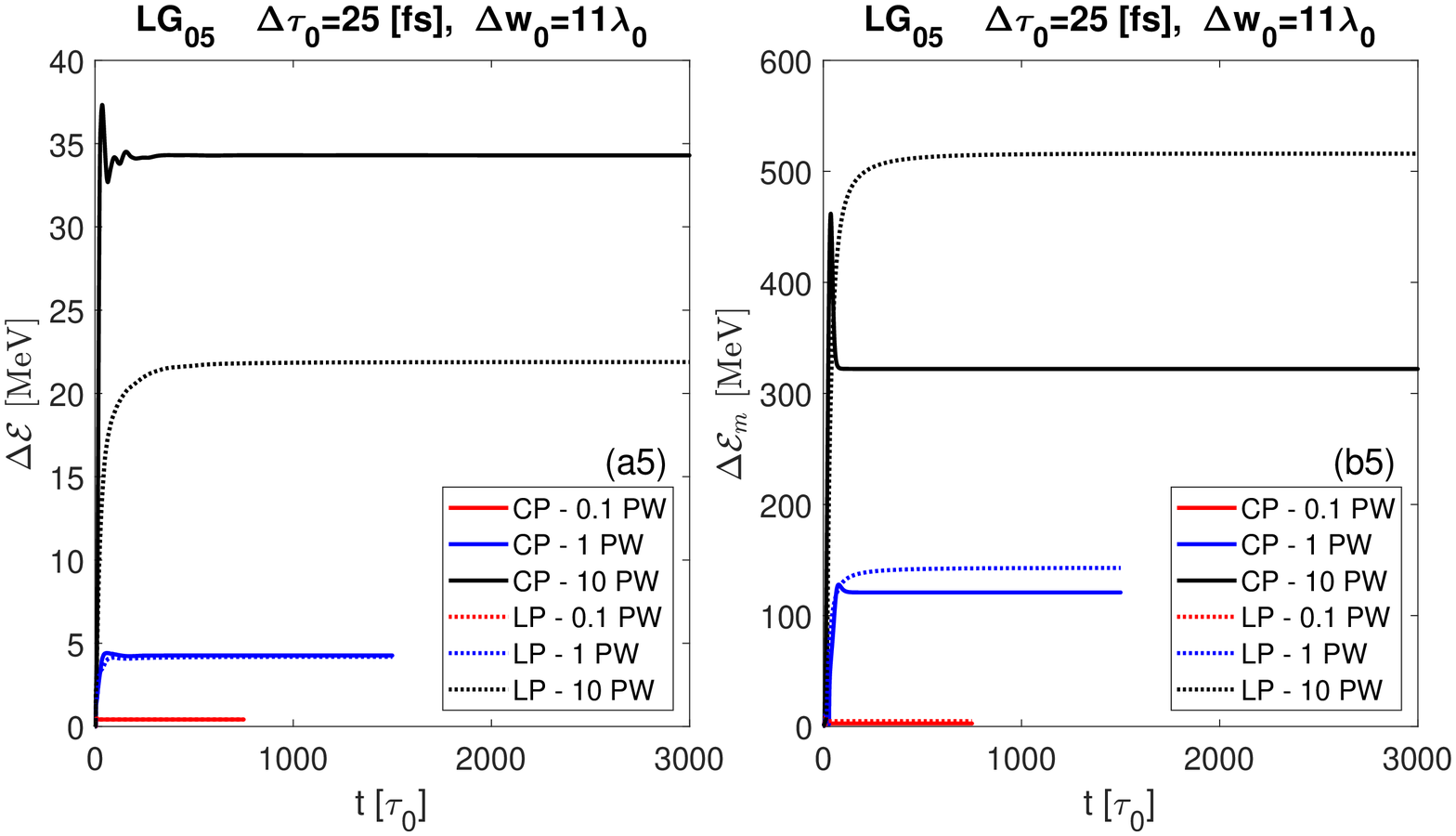}
\caption{Similar to Figs. \ref{fig:Energy_CPLP_LG0x_w06L}, but now all figures correspond to lasers 
with $\Delta \tau_0=25$ fs and $\Delta w_0=11 \lambda_0$ beam waist radius. 
The full and dotted lines correspond to a CP and LP laser  
of $P_0=\left\{0.1, 1, 10\right\}$ PW power, with red, blue and black, respectively.}
\label{fig:Energy_CPLP_LG0x_w011L}
\end{figure*}
\begin{figure*}[hbt!]
\vspace{-0.2cm} 
\includegraphics[width=8.5cm, height=4.4cm]
{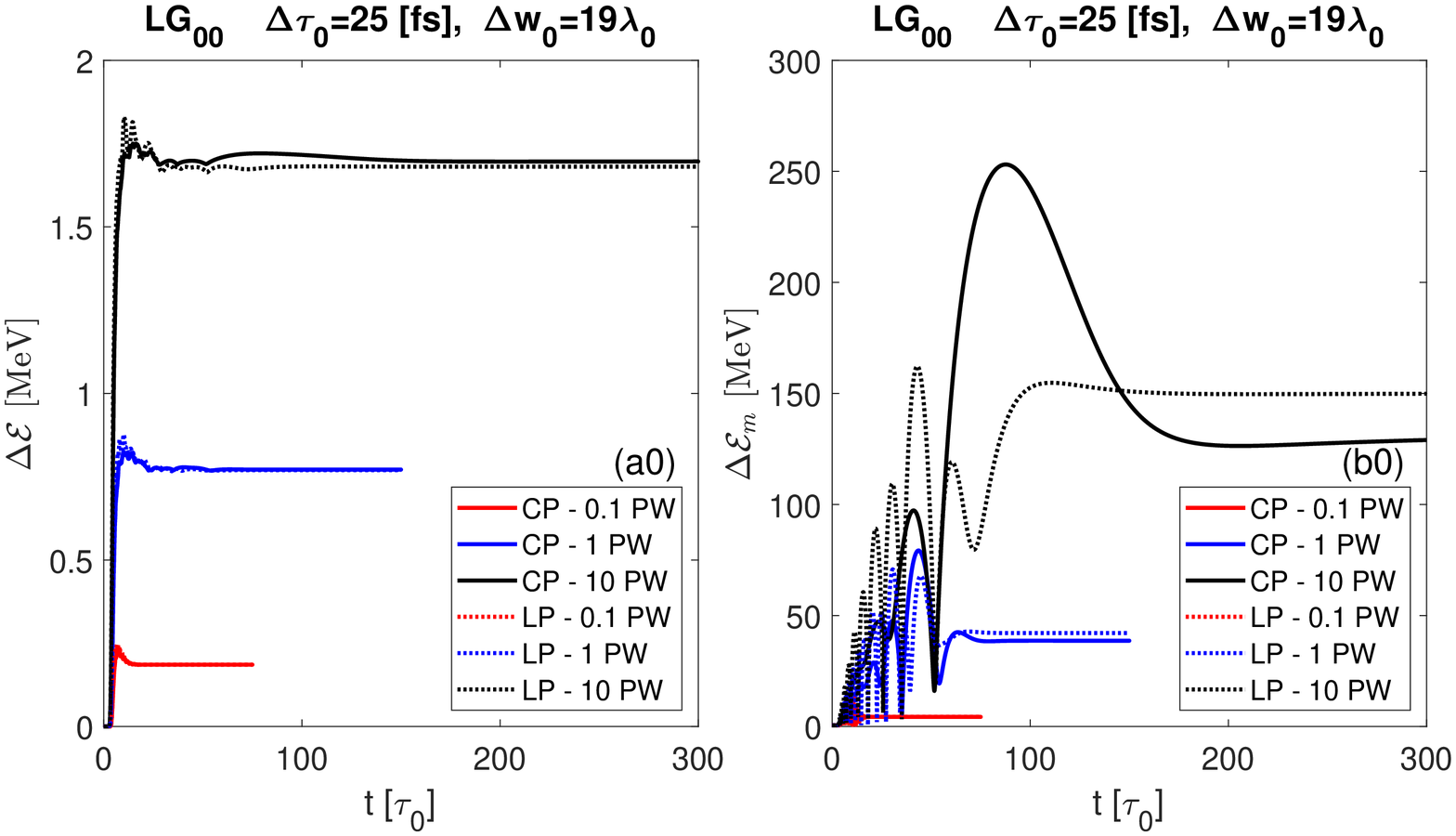} 
\includegraphics[width=8.5cm, height=4.4cm]
{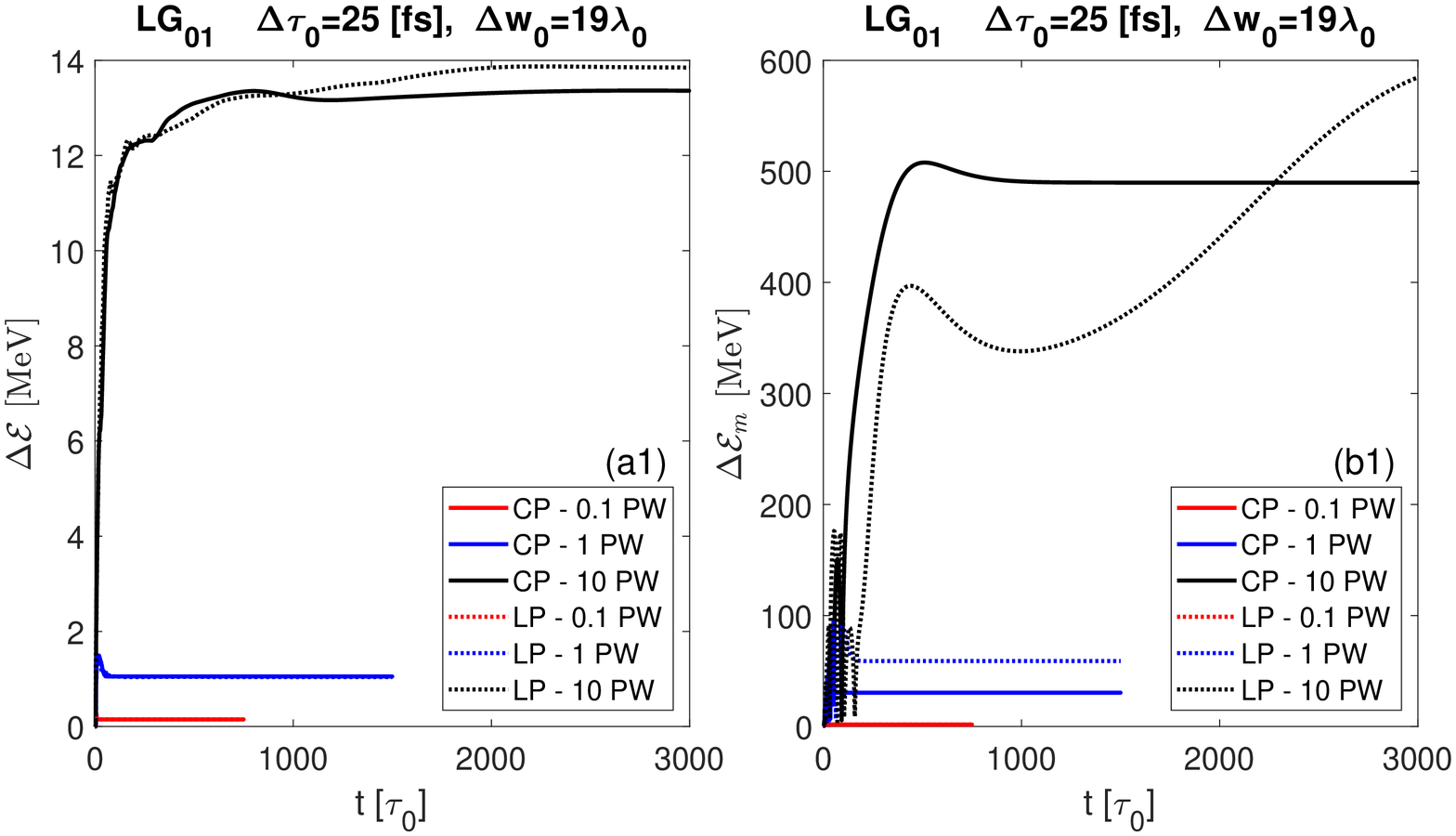} 
\includegraphics[width=8.5cm, height=4.4cm]
{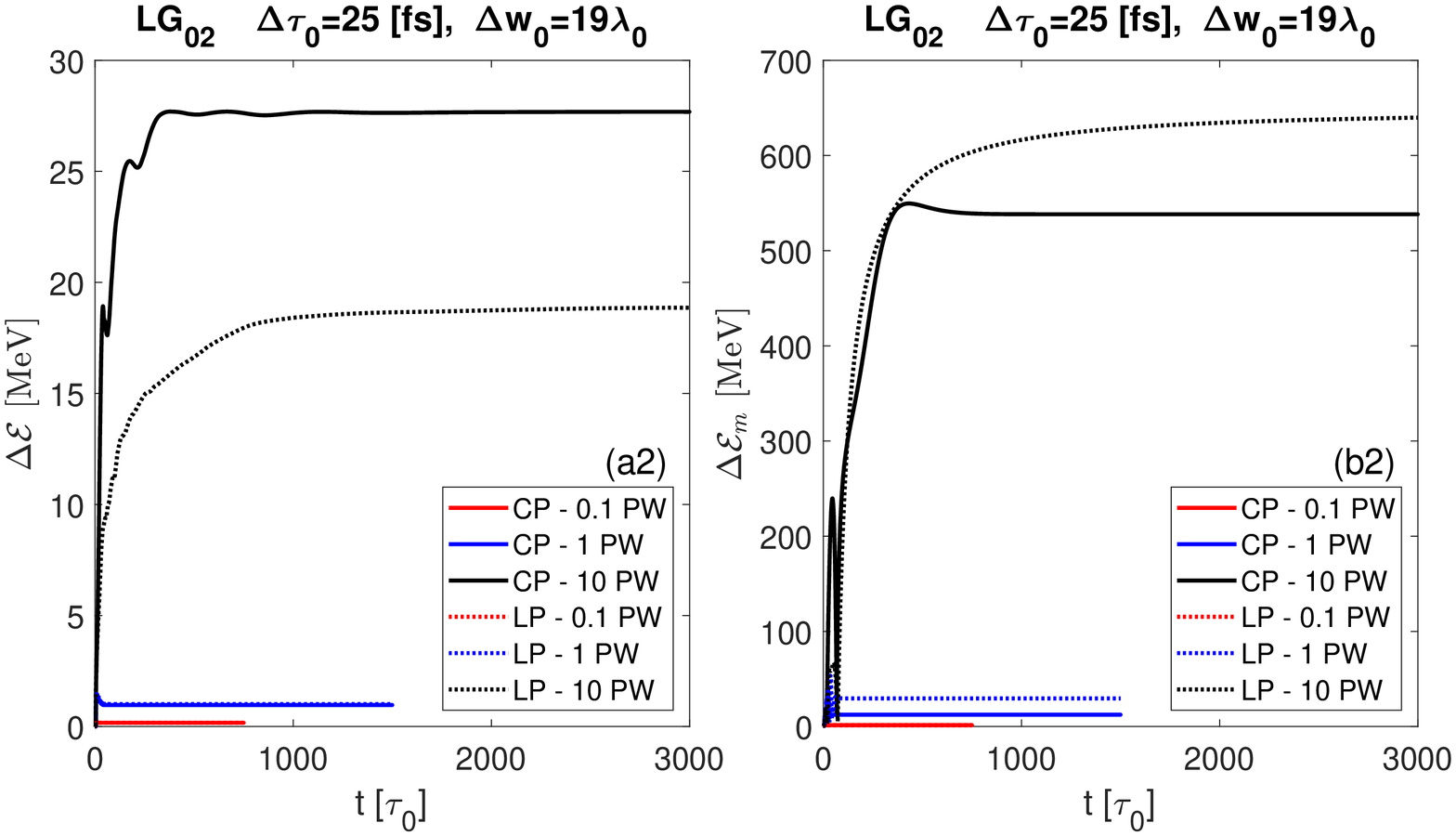} 
\includegraphics[width=8.5cm, height=4.4cm]
{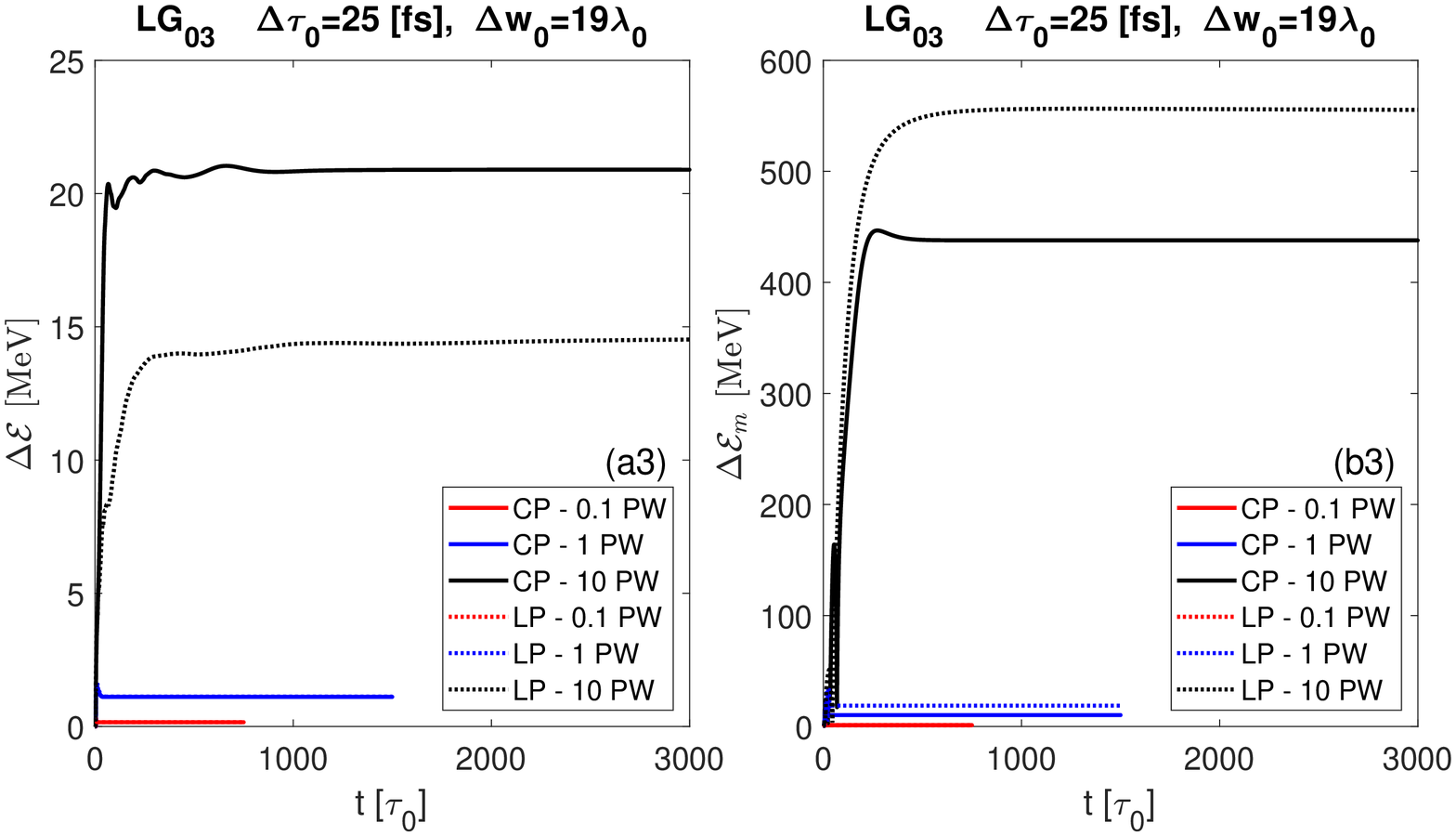}
\includegraphics[width=8.5cm, height=4.4cm]
{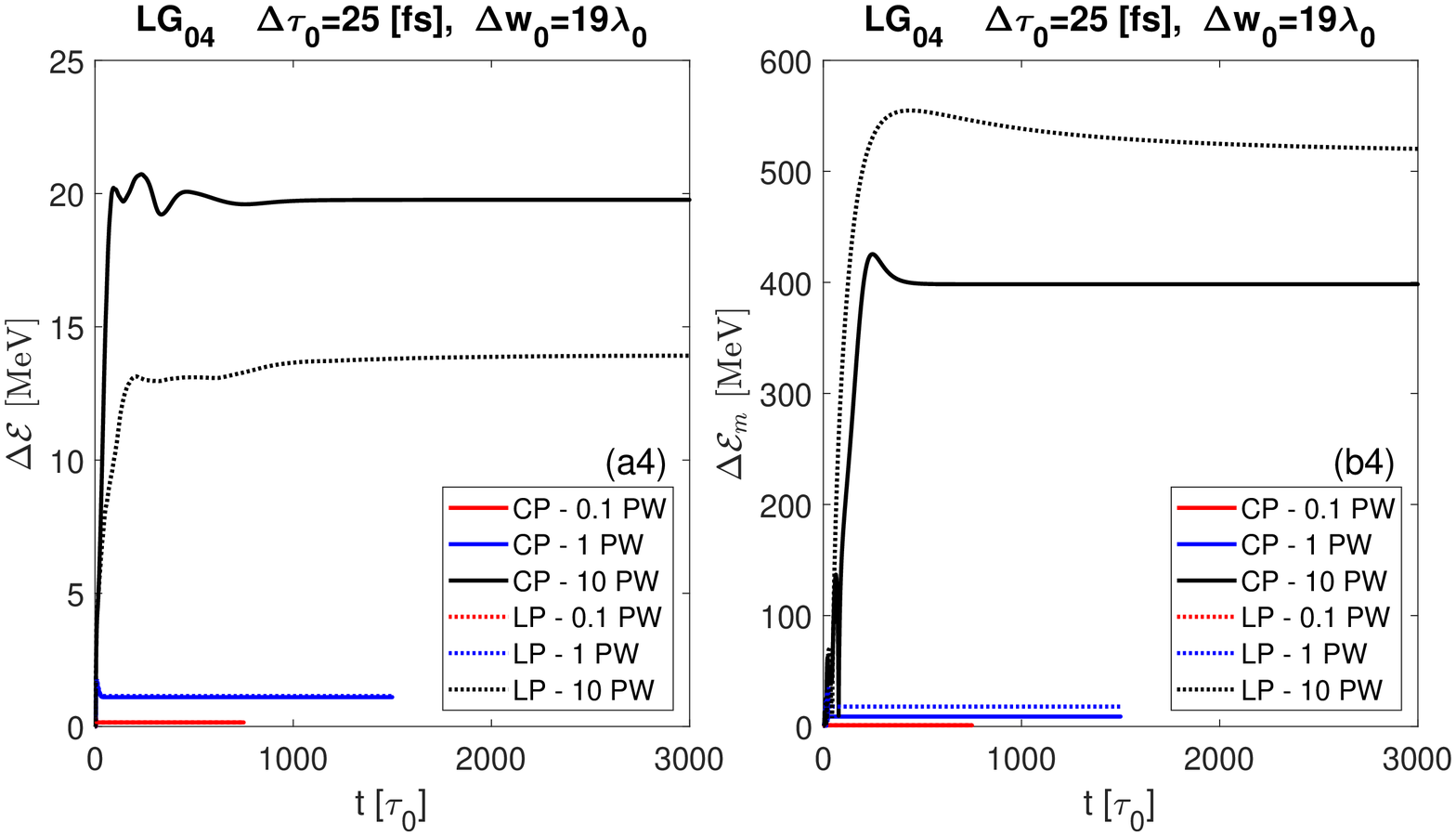}
\includegraphics[width=8.5cm, height=4.4cm]
{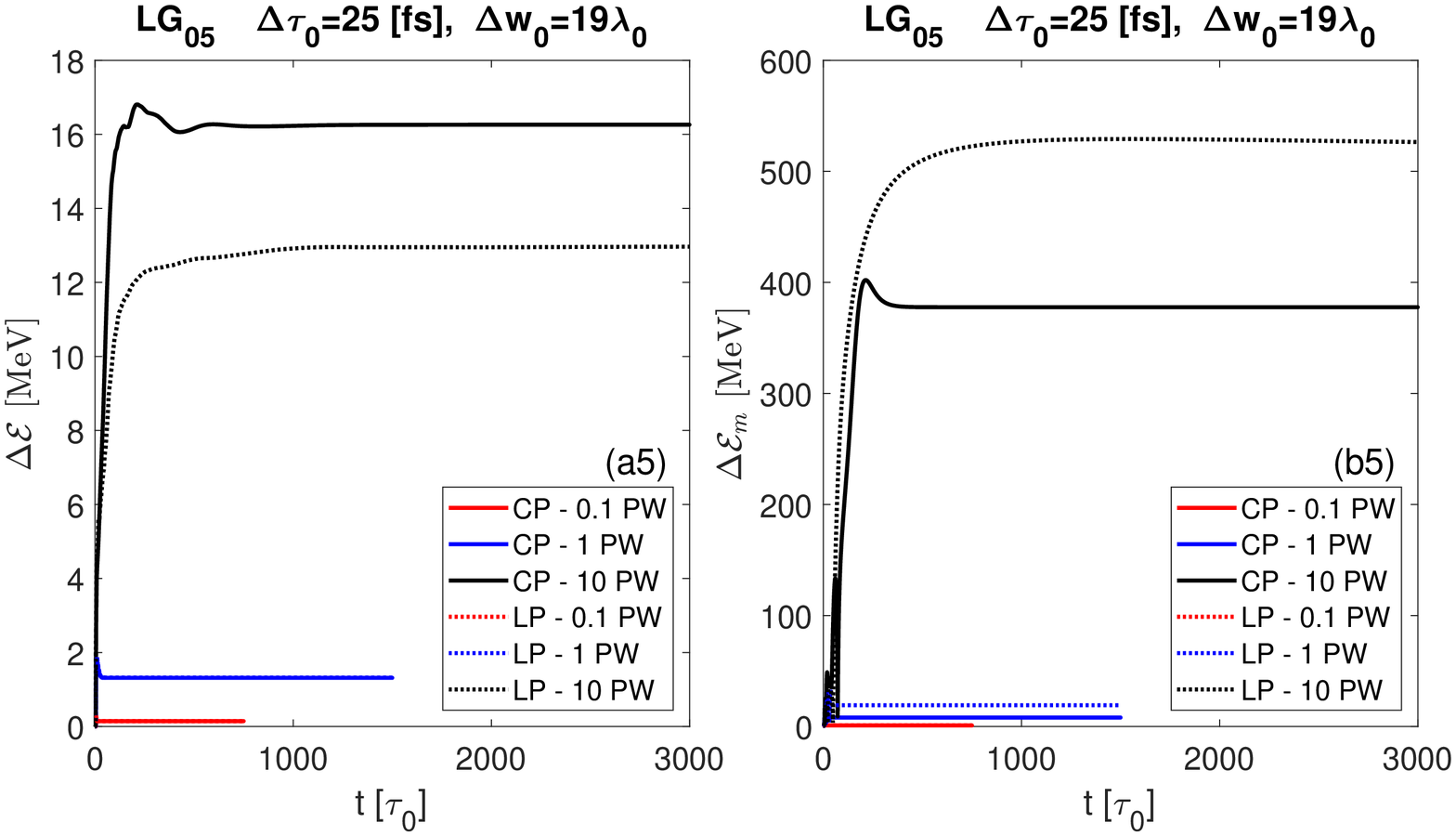}
\caption{Similar to Figs. \ref{fig:Energy_CPLP_LG0x_w06L} and 
Figs. \ref{fig:Energy_CPLP_LG0x_w011L}. 
All figures correspond to lasers with $\Delta \tau_0=25$ fs and $\Delta w_0=19 \lambda_0$ beam waist radius. 
The full and dotted lines correspond to a CP and LP laser  
of $P_0=\left\{0.1, 1, 10\right\}$ PW power, with red, blue and black, respectively.}
\label{fig:Energy_CPLP_LG0x_w019L}
\end{figure*}

\section{Results}
\label{Results}
\subsection{Optimal beam waist}
\label{Optimal_waist}

First of all we are interested in the values of the beam waist for different laser powers 
that lead to maximal energy gains, in case of the Gaussian beam 
$\text{LG}_{00}$, 
and the helical beams $\text{LG}_{01},\text{LG}_{02}$, $\text{LG}_{03}$, 
$\text{LG}_{04}$ and $\text{LG}_{05}$.

The energy gain of electrons interacting with an $\text{LG}_{0m}$ laser pulse is 
a function of the initial location of electrons, the laser spot size and azimuthal mode $m$.
To estimate the value of the beam waist that correspondingly leads 
to maximum energy gains for a given laser power we have varied the initial 
position of electrons uniformly $x_{0,i}=y_{0,i}\equiv \left\{0.05,0.1,0.15,\cdots,2.5 \right\} w_0$ 
at $z_{0,i}=0$,
and calculated the respective energy gain of a single electron for each position,
\begin{equation}
\Delta \mathcal{E}_{i}\left( t\right) \equiv
\mathcal{E}_{i}\left( t\right) - \mathcal{E}_{i}\left( t_{0}\right) 
= \left(\gamma_i - 1\right)m_e c^2 \, ,
\end{equation}
where $\mathcal{E}_i (t_0) =  m_e c^2$.  
Furthermore, for all discrete values of beam waist $\Delta w_0 = \left\{1,2,\cdots,200\right\}\lambda_0$,
we have also calculated the average of these energy gains, 
$\Delta \mathcal{E} = \sum_i^N \Delta \mathcal{E}_i/N$, where $N=50$ corresponds to the number 
of the initial positions of electrons in the transverse plane.

These weighted averages for different laser powers are shown in 
Figs. \ref{fig:w0_scan} for both LP and CP  pulses.
Here the average energy gains
corresponding to laser powers of $P_0=\left\{0.1, 1, 10\right\}$ PW 
are plotted with red, blue and black, as a function of initial spot size. 
The full and dashed lines correspond to CP and LP lasers respectively.

First we discuss these results qualitatively. 
For a given laser power, the amplitude of the electron oscillations along 
the polarization direction increases with intensity. 
Increasing the beam waist decreases the intensity and the scattering of electrons 
decreases hence they remain confined in the pulse being able to gain more energy from the laser, 
until the oscillations become larger than the waist and the electron scatters out from the pulse.

On the other hand increasing the waist of the beam also decreases the longitudinal components 
of the electric field and the Lorentz force, i.e., $E_z$ and $(\vec{v} \times \vec{B})_z$, 
and therefore reduces the net kinetic energy gain of electrons. 
The electrons are accelerated to larger and larger velocities 
in the front part of the pulse and thus the electron trajectories 
are elongated in the direction of the laser propagation while the deceleration 
in the back part of the pulse becomes less efficient.

This means that the beam waist corresponding to the highest average energy gain, 
i.e., the highest peaks of the averages in Figs. \ref{fig:w0_scan}, represent the optimal waist 
for the given laser power and polarization.
For larger power lasers a wider initial beam waist is more optimal to ensure 
that the electrons remain confined inside the pulse to gain more energy.

Therefore using Figs. \ref{fig:w0_scan} we can approximate of the beam waists
corresponding to the peaks in net energy gain. 
In case of the fundamental Gaussian beam,  
the optimal beam waist at FWHM correspond to few tens of wavelengths, 
$\Delta w_0=\left\{13,23,41\right\}\lambda_0$, for laser powers of
$P_0 = \left\{0.1,1,10\right\}$ PW, such that 
the net energy gain increases about $\Delta \mathcal{E}\approx \sqrt{10}$ for every order 
of magnitude increase in laser power. 
Even with those optimal values the average energy gain of only a few MeV was observed in $\text{LG}_{00}$ beams, 
see Ref. \cite{Molnar_2020} for more details. 
Note also that in case of the Gaussian pulse the outcome is independent on the polarization, 
and the averaged results overlap.
This behavior is similar for the $\text{LG}_{01}$ helical beam, but for higher 
modes the distinction between CP and LP pulses becomes more apparent.

Now, averaging the energy gains $\Delta \mathcal{E}$ of the helical beams $\text{LG}_{01}$ and 
$\text{LG}_{02}$ for a circularly polarized laser from Fig. \ref{fig:w0_scan} we have 
approximated the optimal beam waists at FWHM. 
These optimal beam waists are $\Delta w_0=\left\{6,11,19\right\}\lambda_0$ 
corresponding to increasing laser power of $P_0 = \left\{0.1,1,10\right\}$ PW. 
These approximated values represent the chosen optimal values for direct electron 
acceleration in all helical beams of interest, $\text{LG}_{01}, \cdots,\text{LG}_{05}$. 
Note however that the optimal beam waist also reduces slightly as the mode number increases, 
see Fig. \ref{fig:w0_scan}. 
Therefore the previously chosen beam waists are suboptimal for $\text{LG}_{04}$ and $\text{LG}_{05}$ modes 
where even smaller beam waist would be more favorable for larger energy gains.

The most important observation in case of CP beams is that the optimal spot sizes 
of helical beams are more than twice smaller than for the fundamental Gaussian beam. 
In case of LP beams this difference further increases with the mode index 
$\left\vert m\right\vert  \ge 1$.
Furthermore for these relatively tight initial waists,
of a few wavelengths, the helical beams might lead to almost an order of magnitude larger
net energy gains compared to the fundamental Gaussian beam at high laser power. 

A straightforward explanation can be formulated in terms of the ponderomotive
force, $F^{0m}_P = -\nabla \Phi^{0m}$, where the ponderomotive
potential, the cycle-averaged oscillation energy, is directly proportional 
to the intensity, $\Phi^{0m} \simeq e^2 |\vec{E}_{0m}|^2/(4 m_e \omega^2_0)\sim I^{0m}_0$, 
shown in Fig. \ref{fig:LG0x_profiles}.
The ponderomotive force causes the charges oscillating
in an inhomogeneous electric field to drift from where the electric field is
larger to where it is smaller.
Therefore the immediate consequence of the transverse ponderomotive force is 
the scattering of charges from "regions" of higher to lower electric field intensity. 
Due to the fact that the intensity of the helical beams have a wide convex region the
charges found this region are naturally driven to the beam center with zero intensity. 

In other words, the transverse ponderomotive force in Gaussian beams is always positive, 
hence the electrons are scattered outwards from the pulse.
In helical beams the transverse ponderomotive force of the convex region is negative, 
and hence the electrons are effectively trapped inside the "hollow" pulse, see Fig. \ref{fig:LG0x_profiles}.
This leads to less spread, better focusing and collimated electron trajectories confined near the 
axis of propagation. 
These captured electrons are accelerated further by 
the longitudinal ponderomotive force while continue to gain more
energy through the phase synchronization process leading to larger energy gains, 
see Ref. \cite{Akou_2020} for more details.

\subsection{Energy gain for optimal beam waists}
\label{Energy_gain}

Using the previously given initial conditions together with the optimal values listed 
in Table \ref{Power_table} for both CP and LP lasers
of $P_0=\left\{0.1, 1, 10\right\}$ PW power and initial beam waist of 
$\Delta w_0 = \left\{6, 11, 19 \right\} \lambda_0$, we have numerically 
calculated the direct laser driven electron acceleration corresponding 
to the fundamental Gaussian $\text{LG}_{00}$, 
and $\text{LG}_{0m}$ helical beams with azimuthal modes 
$m=\left\{1,2,3,4,5\right\}$. 

In Figs. \ref{fig:Efields_LG0x} we have plotted all Cartesian components of
the electric field as function of time in units of the normalized field intensity $a_0$ 
as mapped by $N_e=6000$ accelerated electrons with lower index-$i$. 
In $\text{LG}_{00}$ beams the electrons are
accelerated at the front part of the pulse, but without reaching the
available peak intensity of the pulse, the electrons are scattered out by the 
transverse ponderomotive force. 
In helical beams, e.g., $\text{LG}_{01}$ and $\text{LG}_{05}$, the available "peak" intensities
are about $2.5$-- and $5.5$--times smaller than in $\text{LG}_{00}$.
However these intensity maxima are located further away from the midpoint, i.e., $r=0$, the location 
of maximum intensity of the Gaussian beam and zero intensity for the helical beams, see Fig. \ref{fig:LG0x_profiles}.
Therefore the electrons found in the convex part of the pulse are captured
and accelerated for a much longer time, being clearly visible on 
Figs. \ref{fig:Efields_LG0x}, where the time axes of the electric field components 
are an order of magnitude longer for the helical beams.
This also means that the trajectories of some electrons captured
by helical beams are up to an order of magnitude longer in the direction of
the laser propagation. Furthermore as apparent in Figs. \ref{fig:Efields_LG0x} 
the electrons also have a twisted circular motion 
with intertwining trajectories, similar as presented in Ref. \cite{Ju_2018}.

To further elucidate these issues, in Figs. \ref{fig:Position_CPLP_LG0x_w011L} we have plotted the net
energy gain of electrons as a function of the initial radial distance, 
$r_{0,i} = \sqrt{x^2_{0,i} + y^2_{0,i}}$, at the origin $z_{i,0}=0$,
for varying laser power. 
Here all figures correspond to an initial beam waist of $\Delta w_0=11\lambda_0$, while CP and LP pulses 
are plotted with "o" and "x" respectively.
Similarly to the previously presented Figures, the red, blue and black correspond 
to $P_0=\left\{0.1, 1, 10\right\}$ PW laser power.
In the 1st row of Figs. \ref{fig:Position_CPLP_LG0x_w011L}: the 1st and 2nd figures correspond to Gaussian beams, 
while the 3rd and 4th are for helical beams $\text{LG}_{01}$.
Similarly in the 2nd and 3rd rows: the helical beams, 
$\text{LG}_{02}$, $\text{LG}_{03}$ and $\text{LG}_{04}$, $\text{LG}_{05}$, 
with different polarizations are shown.

Here we observe that the net-energy gain as a function of the initial radial 
distance also reflects the initial intensity profiles shown on Fig. \ref{fig:LG0x_profiles}. 
The intensity of Gaussian beam falls off exponentially as a function of the radius, hence 
the electric charges found further away form the center gain less and less energy. 
In the case of the helical beams the largest acceleration happens within the convex part
of the intensity curves. 
Furthermore as the distance between the center and the intensity peak widens with the azimuthal index, 
the peaks in intensity of the $\text{LG}_{0m}$ helical beams 
is also decreasing, see Fig. \ref{fig:LG0x_profiles}. 
This explains why there is less and less net energy gained around the middle of the
helical beams with increasing mode index $m$, see 
Fig. \ref{fig:Position_CPLP_LG0x_w011L}.

In Figs. \ref{fig:Histogram_CPLP_LG0x_w011L} the histograms of the energy 
$\Delta \mathcal{E}_i$ 
and the polar angle $\phi_i$ of electrons is shown, after the interaction with the pulse.
Here the 3-dimensional polar angle is, 
$\phi_i = \arccos \left(z_i/R_i\right)$, where 
$R_{i} = \sqrt{x^2_{i} + y^2_{i} + z^2_{i}}$ is the radial distance from the origin.
All histograms correspond to a CP laser of $P_0 = 10$ PW power and 
$\Delta w_0=11\lambda_0$ spot. 
In Figs. \ref{fig:Histogram_CPLP_LG0x_w011L} the 1st row from left to right, (a0), (b0) and (a1), (b1), 
shows the histograms of energy and polar angle corresponding to $\text{LG}_{00}$ and $\text{LG}_{01}$ beams.
Similarly, the 2nd and 3rd rows: figures (a2), (b2), (a3), (b3), and (a4), (b4), (a5), (b5), 
show the outcome from the helical beams
$\text{LG}_{02}$, $\text{LG}_{03}$, and $\text{LG}_{04}$, $\text{LG}_{05}$.

These histograms once again reflect the difference in the ponderomotive force and its influence on 
the energy gain as well as the angular distribution of electrons.
Due to the interaction with pulse the electrons have scattered out with $\phi_i$ polar angle.
In case of the fundamental Gaussian beam this polar angle is predominantly in the direction orthogonal to 
the direction of laser propagation, i.e., $\phi_i \geq 80$ degrees.
For helical beams with increasing azimuthal mode we observe an increasing 
number of electrons that are scattered parallel to the longitudinal axis, 
at polar angles of $0 \leq \phi_i < 15$ degree, and at the same time fewer electrons 
in the orthogonal directions.
This obviously means that helical beams lead to a larger number of collimated electrons than fundamental 
Gaussian beams, while their number is also increasing with increasing mode index.

Furthermore, the distribution of energy gain is also different in helical beams.
The electrons are distributed following a multimodal distribution, 
leading to at least a second peak at finite energy
at about $150$ MeV energy for the helical beams, $\text{LG}_{03}$, $\text{LG}_{04}$ and $\text{LG}_{05}$.
Although here we have only shown the results for $P_0=10$ PW laser, a very similar behavior is observed for the 
$P_0=1$ PW laser, with a peak at about $50$ MeV energy.
For a $P_0=0.1$ PW laser, the intensities of the helical beams are so weak, 
that these striking differences disappear.

In Figs. \ref{fig:Energy_CPLP_LG0x_w06L}, \ref{fig:Energy_CPLP_LG0x_w011L} 
and \ref{fig:Energy_CPLP_LG0x_w019L} corresponding to initial beam waists 
of $\Delta w_0=6\lambda_0$, $\Delta w_0=11\lambda_0$ and $\Delta w_0=19\lambda_0$, 
we have plotted the average net energy and the 
largest energy gained, $\Delta \mathcal{E}_m$, both from circularly and linearly polarized $\text{LG}_{0m}$ pulses. 
Here the average energy is calculated from $\Delta \mathcal{E}= \frac{1}{N_e} \sum_{i=1}^{N_e} \Delta \mathcal{E}_i$, 
where $N_e = 6000$ is the number of electrons. These averages are shown with full lines for circularly polarized and with dotted lines for a linearly polarized lasers.
Similarly as before, red, blue and black, respectively represent the 
power of the laser $P_0=\left\{0.1, 1, 10\right\}$ PW.
In the 1st row of Figs. \ref{fig:Energy_CPLP_LG0x_w06L}, \ref{fig:Energy_CPLP_LG0x_w011L}, 
\ref{fig:Energy_CPLP_LG0x_w019L}: the 1st and 2nd figures, (a0) and (b0), show the average energy gained within 
the fundamental Gaussian beams while the 3rd and 4th figures, (a1) and (b1), show the same for 
the helical beams  $\text{LG}_{01}$ with given waist.
In the 2nd row: the 1st and 2nd figures, (a2) and (b2), correspond to $\text{LG}_{02}$, while
the 3rd and 4th figures, (a3) and (b3), correspond to $\text{LG}_{03}$.
Similarly, the 3rd row: the figures (a4), (b4) and (a5), (b5), presents the outcome 
for $\text{LG}_{04}$ and $\text{LG}_{05}$ helical beams.

We observe that the mean energy gain of electrons for any given laser power is largest for an optimal waist, 
and this optimal waist is increasing with increasing laser power.
Therefore for a $P_0=0.1$ PW laser the optimal waist size is $\Delta w_0 = 6 \lambda_0$, 
compare the mean energies, i.e., the red lines, in Figs. \ref{fig:Energy_CPLP_LG0x_w06L}, 
\ref{fig:Energy_CPLP_LG0x_w011L} and \ref{fig:Energy_CPLP_LG0x_w019L}.
Similarly by comparing the blue lines, for a $P_0=1$ PW laser the waist size that leads 
to the largest energy gains is $\Delta w_0 = 11 \lambda_0$, while for a $P_0=10$ PW laser 
the optimal waist size is $\Delta w_0 = 19 \lambda_0$, in accordance with the results of Sect. \ref{Optimal_waist}.

On the other hand we also recognize that the average energy gain in helical beams is
increasing with increasing azimuthal index $\left\vert m\right\vert \ge 1$. 
This is evident by comparing the blue and black lines on Figs. \ref{fig:Energy_CPLP_LG0x_w06L}, 
\ref{fig:Energy_CPLP_LG0x_w011L} and Figs. \ref{fig:Energy_CPLP_LG0x_w019L}. 
For a larger initial waist there is larger energy gain for higher laser power.
However, in Figs. \ref{fig:Energy_CPLP_LG0x_w019L} we also observe that for the 
largest waist and largest laser power, the average energy is increasing up to 
$\left\vert m\right\vert \leq 3$ while it is 
decreasing for higher modes. This was somewhat expected based on Figs. \ref{fig:w0_scan}, 
meaning that for the highest power lasers with $\left\vert m\right\vert \geq 3$ modes the optimal 
waist should be smaller than $\Delta w_0 \leq 19 \lambda_0$. 

The effect of laser polarization on energy gain is a delicate matter as already suggested by Figs. \ref{fig:w0_scan}.
The average net energy gain in fundamental Gaussian beams is largely independent on the polarization.
However for helical beams the outcome is different in each case depending on the beam waist 
and mode index and therefore the difference is from a few percent to up to $50 \%$. 
We have also observed that a larger average energy gain slightly favors the circularly polarized 
helical beams with mode indices $\left\vert m\right\vert \geq 3$ for any beam waist and laser power.
For linearly polarized helical beams the optimal waist size should be slightly smaller than 
in the circularly polarized case to obtain similar energy gains, as can be concluded from the results.

The average net energy gain in Gaussian pulses, for the highest laser power and 
largest waist is less than $2$ MeV, while the highest energy electrons are  
$\mathcal{E}_{max} \approx 150$ MeV, see Fig. \ref{fig:Energy_CPLP_LG0x_w019L}.
As previously discussed, the energy gains in fundamental Gaussian pulses may be a little higher 
for optimal beam waists Ref. \cite{Molnar_2020},
but even so these values are easily surpassed in helical beams leading to an average of about a few and up to $45$ MeV, 
and in some cases with highest energy electrons of $\mathcal{E}_{max} \approx 650$ MeV or more.
Such energy gains are still below than the so-called ponderomotive limit
$\Delta \mathcal{E} \approx m_e c^2 a^2_0/2$ \cite{Stupakov_2001,Dodin_2003}, 
where the intensity maxima of higher-order helical beams can be more than $5$-times lower than of Gaussian beams.

\section{Conclusions}
\label{Conclusions}

In this paper we have studied and compared the direct laser-driven electron acceleration in Laguerre-Gaussian beams 
with azimuthal mode indices, $\text{LG}_{00}$ corresponding to the fundamental Gaussian beam, 
and helical beams $\text{LG}_{01}$, $\text{LG}_{02}$, $\text{LG}_{03}$, $\text{LG}_{04}$ 
and $\text{LG}_{05}$.

We have found that the acceleration of electrons from rest is vastly different in
helical beams compared to the fundamental Gaussian beam, mainly due to the difference 
in ponderomotive forces.
Most importantly for relatively tight initial waists, the helical beams lead 
to at least an order of magnitude larger energy gains compared to the fundamental 
Gaussian beam, and this energy gain also increases with the azimuthal mode 
$\left\vert m\right\vert \geq 1$.

For laser powers of $P_0 = \left\{0.1,1,10\right\}$ PW, the optimal waist of helical beams 
leading to the most energetic electrons are more than twice smaller, $\Delta w_0=\left\{6,11,19\right\}\lambda_0$ 
than in case of the fundamental Gaussian beam \cite{Molnar_2020}.
This also means that the beam waist that is optimal for helical beams is suboptimal for Gaussian 
beam and vice versa.

Finally, compared to Gaussian beams the electron trajectories in helical beams 
are confined in the direction of laser propagation leading to collimated electrons with trajectories 
that are at least an order of magnitude longer in the direction of the laser propagation.
These conclusion hold for both linearly and circularly polarized lasers.

\begin{acknowledgments}
The authors are thankful to S. Ataman, D. Doria, J. F. Ong, K. Tanaka and S. Tzenov 
as well as the referees for corrections and suggestions.
E.~Moln\'ar thanks H. S. Ghotra for the comparison of early results 
and valuable discussions.
D. Stutman acknowledges support by a grant of 
Ministry of Education and Research, CNCS-UEFISCDI, project number 
PN-IIIP4-ID-PCCF-2016–0164, within PNCDI III.
The authors are thankful for financial support from the Nucleu Project PN 19060105. 
\end{acknowledgments}


\end{document}